\documentclass[aps,physrev,reprint,amsmath,amssymb,superscriptaddress,floatfix]{revtex4-2}

\usepackage{graphicx}
\usepackage{dcolumn}
\usepackage{bm}
\usepackage{amsfonts} 
\usepackage{CJK}
\usepackage{subfigure}
\usepackage[dvipsnames]{xcolor}
\usepackage[colorlinks,
linkcolor=NavyBlue,
anchorcolor=blue,
urlcolor=NavyBlue,
citecolor=NavyBlue
]{hyperref}    
\usepackage{marvosym}
\usepackage{mathtools}
\usepackage{placeins}
\usepackage{enumerate}
\usepackage{threeparttable}
\usepackage{booktabs}
\usepackage{csquotes}

\providecommand{\codexnew}[1]{#1}
\providecommand{\codexnewcolor}{}
\providecommand{\codexflag}[1]{#1}
\colorlet{orange}{black}

\begin{document}

\preprint{APS/123-QED}

\title{Dynamics Creation through Neural Dynamical Transfer Learning}%

\author{He Ma}
\thanks{These authors contributed equally to this work.}
\affiliation{School of Mathematical Sciences, SCMS, and SCAM, Fudan University, Shanghai 200433, China}
\affiliation{Research Institute of Intelligent Complex Systems, Fudan University, Shanghai 200433, China}

\author{Qiyang Ge}
\thanks{These authors contributed equally to this work.}
\affiliation{Research Institute of Intelligent Complex Systems, Fudan University, Shanghai 200433, China}

\author{Yu Meng}
\affiliation{Research Institute of Intelligent Complex Systems, Fudan University, Shanghai 200433, China}

\author{Celso Grebogi}
\affiliation{Institute for Complex Systems and Mathematical Biology, King's College,
University of Aberdeen, Aberdeen AB24 3UE, United Kingdom}

\author{Wei Lin}
\email{wlin@fudan.edu.cn}
\thanks{Corresponding author.}
\affiliation{School of Mathematical Sciences, SCMS, and SCAM, Fudan University, Shanghai 200433, China}
\affiliation{Research Institute of Intelligent Complex Systems, Fudan University, Shanghai 200433, China}
\affiliation{Shanghai Artificial Intelligence Laboratory, Shanghai 200232, China}
\affiliation{MOE Frontiers Center for Brain Science and State Key Laboratory of Medical Neurobiology,
Fudan University, Shanghai 200032, China}




 \date{\today}


\begin{abstract}
Data-driven machine learning has established a robust foundation for reconstructing nonlinear dynamical systems from observations, primarily for the purposes of forecasting and control. However, most existing efforts focus on recovering specific observed dynamics rather than the generative synthesis of new ones. Inspired by image fusion and style transfer, we introduce a neural network framework termed Neural Dynamical Transfer Learning (NDTL) to create new systems with prescribed dynamics from pairs of parent nonlinear dynamical systems. By computing fundamental dynamical signatures, including the intrinsic dimension, the Kaplan-Yorke dimension, the invariant measure statistics, and the Lyapunov spectrum, we demonstrate that NDTL preserves key features inherited from the parent models while simultaneously generating novel dynamics. \codexnew{Beyond these validation examples, NDTL induces a criterion for dynamics classification, creates stable oscillatory coexistence in the Hastings-Powell food chain model, produces interpretable epidemiological models, and provides a chaotic source for image encryption.}
\end{abstract}

\maketitle

\date{\today}
 
\section{Introduction}
In the era of big data, data-driven methodologies have become increasingly pivotal for discovering first principles and modeling dynamical systems across diverse scientific fields. These methods range from classical statistical and dynamical methods such as delay-embedding~\cite{takens2006}, data mining~\cite{kruth2009EL}, and causal inference~\cite{ccm2012science} to model-based methods grounded in physical laws and parameter estimation~\cite{celso_sindy2011PRL,celso_sindy2011PRX,dmd2010JFM}, and further to machine learning methods~\cite{rc_prediction_chaos2018PRL, liqiaofeng_metalearning2023PRL, ml4nonlinearpde2018jmlr, data_driven_pde2017SA}.

Among these methodologies, machine learning approaches based on neural networks have emerged as powerful tools that often outperform traditional statistical and model-based methods. These approaches are broadly categorized by whether they incorporate explicit dynamical priors. On the one hand, purely data-driven architectures such as recurrent neural networks, including reservoir computing~\cite{rc_prediction_chaos2018PRL, rc_review_sun_lin2024nc, highoder_rc2024nc} and gated variants like GRUs and LSTMs~\cite{lstm_review2019}, as well as attention-based transformer models~\cite{transformer2017nips, transformer4fluid2024nc}, have been widely used to learn complex temporal patterns in nonlinear systems. On the other hand, physics-informed models such as physics-informed neural networks \cite{pinn2021nrp, pinn_review2019jcp, hnn_pre2020, zjd_koopman2024PRR, lu2026interpretable} and neural ordinary differential equations \cite{node2018nips, ai_pontryagin2022nc, ndde2021iclr} embed physical laws into the learning process, enabling interpretable modeling and robust extrapolation. Collectively, these architectures have demonstrated remarkable success in recovering governing structures, reconstructing dynamics, forecasting chaos, and generalizing across diverse nonlinear dynamical regimes.

Beyond these advances, there has been growing interest in applying generative machine learning to dynamical systems~\cite{generative4ds_gilpin2024}. Recent studies have utilized variational autoencoders and diffusion models to facilitate system reconstruction~\cite{gled_gaoheng_nc2024}, identification~\cite{generative4ds_review_kutz2024}, and even zero-shot forecasting~\cite{panda_gilpin2025}. However, creating new dynamical systems that inherit prescribed behaviors from parent systems while manifesting new scientifically meaningful structures remains largely unexplored. In contrast, the field of computer vision routinely embraces creation through image fusion and generation techniques, such as neural style transfer, which synthesize novel images by blending the content and style of multiple source images~\cite{style_transfer2016cvpr, neural_style_transfer_review2019ieee}. A notable example is a generated painting of the Mona Lisa rendered in a Cubist aesthetic, which successfully fuses the content of the Mona Lisa with the style of a Picasso painting~\cite{style_transfer2016cvpr}.

Inspired by the success of generating visually intelligible paintings, we propose a framework, termed Neural Dynamical Transfer Learning (NDTL), which treats the vector field of a dynamical system as a high-dimensional image and transfers the feature of one parent system onto another. 
This enables the creation of dynamical systems that not only inherit structural features such as attractor geometry or chaotic patterns from their parents, but also possess new structures of scientific significance in their dynamical equations.

\textcolor{orange}{Specifically, NDTL constructs content and style representations from vector fields and examines the dynamical effects of matching them. In Lorenz-like systems, content fusion primarily changes attractor geometry, whereas style transfer can induce or suppress chaos in the examples studied. In addition to system creation, we show that the dynamical signature landscapes generated by NDTL fusion induce a principled criterion for dynamics classification. Finally, applications to ecology, epidemiology, and chaotic image encryption demonstrate the broader utility of NDTL.}

\codexnew{The remainder of this Article is organized as follows. Section~\ref{sec:framework} introduces the NDTL framework. Section~\ref{sec:interpretation} interprets and validates the learned features through analytical examples, controlled fusion, and perturbation experiments. Section~\ref{sec:organization} develops the NDTL-induced classification and examines its consistency across clustering schemes. Section~\ref{sec:applications} presents applications of NDTL, followed by concluding remarks in Sec.~\ref{sec:discussion}.}
 \section{Neural Dynamical Transfer Learning Framework}
\label{sec:framework}
\begin{figure}[t]
\centering
\includegraphics[width=0.48\textwidth]{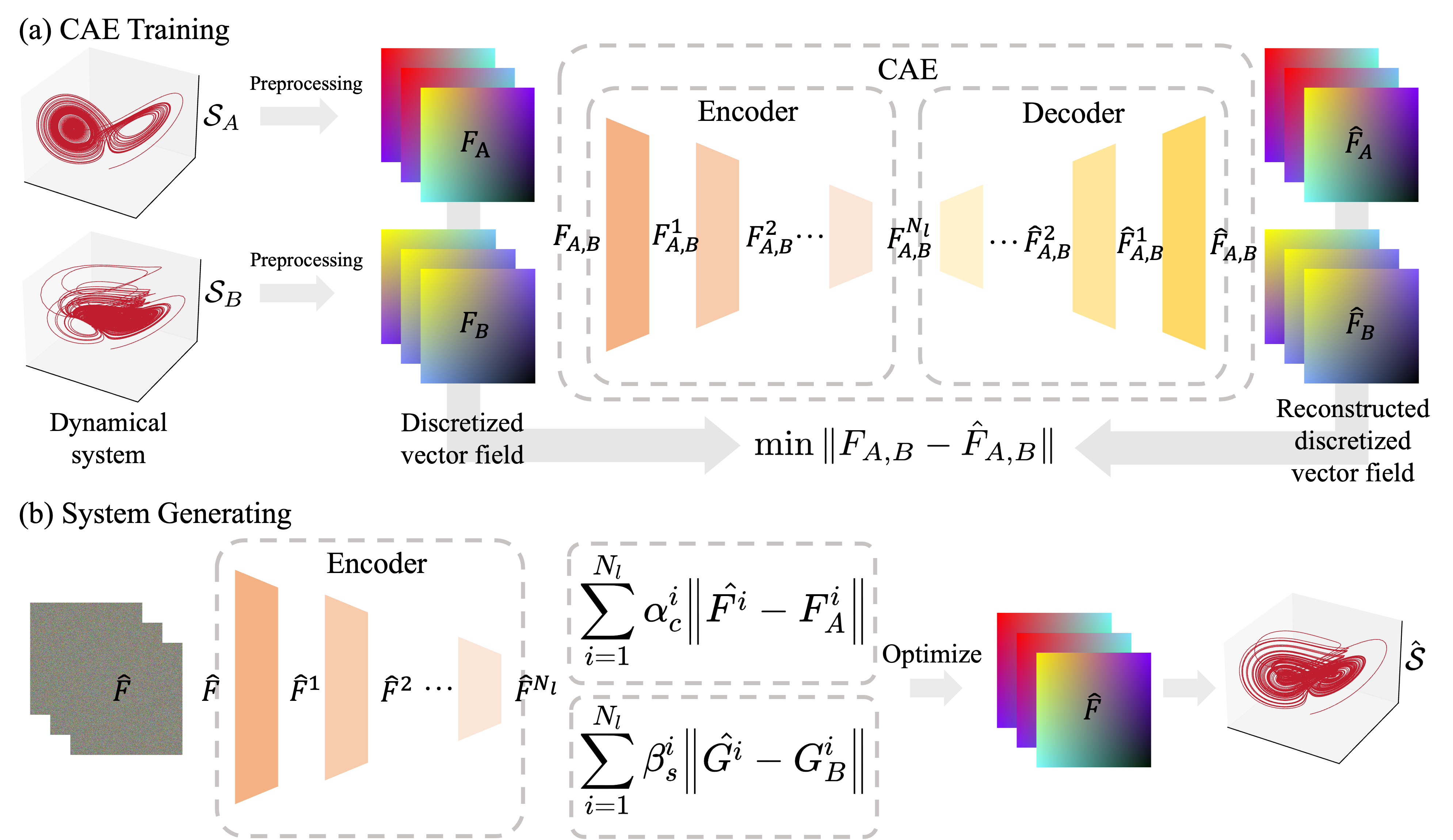}
\caption{Schematic depiction of the NDTL framework. (a) Training procedure of the convolutional autoencoder (CAE) for feature extraction. (b) Optimization process utilizing the trained feature extractor to fuse systems by refining an initially randomized vector field.}
\label{framework}
\end{figure}

To begin with, we consider two $d$-dimensional dynamical systems: \( \mathcal{S}_{A}: \dot{\textbf{x}} =\textbf{f}_{A}(\textbf{x})\) and \( \mathcal{S}_{B}: \dot{\textbf{y}} =\textbf{f}_{B}(\textbf{y})\). Our objective is to fuse the style feature extracted from system \(\mathcal{S}_{B}\) with the content feature extracted from system \(\mathcal{S}_{A}\), as schematically illustrated in Fig.~\ref{framework}. To achieve this, we first discretize their corresponding vector fields \( \textbf{f}_{A} \) and \( \textbf{f}_{B} \) over a common spatial grid, evaluating each vector field at \( N \) discrete grid points. The resulting \( d \)-dimensional vectors are then concatenated into tensors \( F_{A} \) and \( F_{B} \), each of size \( d \times N \). 

From a practical perspective, directly employing high-dimensional tensors as dynamical features is computationally expensive and prone to overfitting, which often results in the created systems that merely replicate the parent dynamics without manifesting novel structure. To mitigate these issues, we utilize compact latent features. High-dimensional tensor data typically reside near a low-dimensional manifold embedded within the original space~\cite{manifold_hypothesis2016jams, bengio_rnn_encoder_decoder2014}. Inspired by recent advances in deep learning for identifying such latent manifold structures, we employ a convolutional autoencoder (CAE) to extract compact latent features from the vector fields~\cite{verbose_cae2011icann}.
 The CAE consists of \(\mathcal{I}\), an encoder, and \(\mathcal{D}\), a decoder.  Specifically,  \(\mathcal{I}\) maps high-dimensional input tensors into a lower-dimensional latent space through a sequence of nonlinear transformations, while \(\mathcal{D}\) reconstructs the original tensors from these compact features.  The network is trained by minimizing the mean squared error (MSE) between the original and reconstructed tensors via back propagation.

\codexnew{Following the training phase, we apply the encoder \(\mathcal{I}\) to each vector field tensor $F_{A,B}$ and denote its activation at the \(i\)-th layer by \(F_{A,B}^{i}\in\mathbb{R}^{c_i\times N_i}\), where $i=1,\dots,L_{l}$. We use these activations as the content features. The style features at the same layer are defined by \(G_{A,B}^{i} = \frac{1}{N_i} F^{i}_{A,B}\big(F^{i}_{A,B}\big)^{\top}\). These operational definitions specify the constraints used in the fusion objective; their dynamical interpretation is examined in Sec.~\ref{sec:interpretation}.}

With these latent features, we formulate the fusion procedure as an optimization problem. Specifically, we initialize a \( d \times N \) tensor \(\hat{F}\) representing the created system's vector field and guide its creation by aligning latent content features from system \(\mathcal{S}_A\) with style features from system \(\mathcal{S}_B\). We define the objective function as
$
\mathcal{L}_{cs}(\hat{F}) = \sum_{i=1}^{L_l}\left(\alpha_{c}^{i}\Vert \hat{F}^{i}-F_{A}^{i}\Vert + \beta_{s}^{i}\Vert \hat{G}^{i}-G_{B}^{i}\Vert\right),
$
where \(\alpha_{c}^{i}\) and \(\beta_{s}^{i}\) are the weights assigned to content and style features of the \(i\)-th layer, respectively. Minimizing \(\mathcal{L}_{cs}(\hat{F})\) via gradient descent ensures that the fused vector field \(\hat{F}\) inherits desired dynamical behaviors from the parent systems. In practice, 
the NDTL extends naturally to more complex fusion architectures by modifying the objective or tuning the relative weights. For instance, setting the loss to 
$\mathcal{L}_{ccs}(\hat{F}) = \sum_{i=1}^{L_l}\left(\alpha_{c}^{i}\Vert \hat{F}^{i}-F_{A}^{i}\Vert + \beta_{c}^{i}\Vert \hat{F}^{i}-F_{B}^{i}\Vert+ \gamma_{s}^{i}\Vert \hat{G}^{i}-G_{C}^{i}\Vert\right)$ enables a fusion that integrates the content of systems A and B with the style of system C (see Appendix~\ref{app:architecture} for implementation details of the CAE model and the NDTL framework).
Moreover, regression techniques~\cite{celso_sindy2011PRL, celso_sindy2011PRX} are employed to extract interpretable governing equations from the fused vector field \(\hat{F}\). Candidate basis functions are selected based on the dynamics of the target system, and various regression methods, including linear regression and LASSO,
are then applied to the discretized vector field (see Appendix~\ref{app:numerics}).

\codexnew{The complete CAE architecture, training protocol, and coordinate and speed normalizations are provided in Appendix~\ref{app:architecture}, while the closed-form regression procedures and numerical evaluation of the dynamical signatures are given in Appendix~\ref{app:numerics}. Below, we examine the dynamical information encoded by these content and style features.}
 
\section{\codexnew{Interpretation and Validation of Content and Style Features}}
\label{sec:interpretation}
\codexnew{The content and style features used in the fusion objective characterize different properties of the vector field. We first interpret these features from their construction and then examine their dynamical roles through a Hamiltonian example, transformation tests, controlled Lorenz-like fusions, and layerwise perturbations.}
\subsection{\codexnew{Content and Style Features of Vector Fields}}
\label{sec:feature_meaning}
\codexnew{We first consider the content feature at the input layer. The tensor \(F\) captures the local, pointwise structure of the vector field. It specifies the local direction of system evolution at the sampled grid points and, provided the discretization is sufficiently dense, encapsulates approximations of the Jacobian matrix at these points. At encoder layer \(i\), the content feature \(F^i\) is the corresponding activation tensor. As the network depth increases, \(F^i\) aggregates vector field structure over increasingly broad receptive fields through successive nonlinear transformations. Content matching therefore constrains spatially resolved local or aggregated vector field structure, depending on the chosen layer.}

\codexnew{Correspondingly, the style feature \(G^i=(1/N_i)F^i(F^i)^\top\) encapsulates the second-order statistics of the feature channels, quantifying correlations among vector field components and nonlinear spatial features. Because the spatial index is contracted in forming \(G^i\), this representation summarizes channel correlations without retaining their exact spatial arrangement.}

\subsection{\codexnew{A Hamiltonian Example}}
\codexnew{A two-dimensional Hamiltonian system provides a concrete illustration of content and style at the input layer.} Consider the Hamiltonian
$H(x_1, x_2)=\frac{1}{2}x_2^2 + V(x_1)$ with $V(x_1)=\frac{1}{2}{x_1^2}-\frac{1}{4}{x_1^4}. $
The equations of motion are given by $\dot{x}_{1}=\frac{\partial H}{\partial x_2}=x_2$ and $\dot{x}_{2}=-\frac{\partial H}{\partial x_1}=-x_1+x_1^3.$
Evaluating these equations at \( N \) discrete grid points $\{(x_1[i], x_2[i])\}$ for \( i \in \{1,2,\cdots, N\} \), we discretize the continuous vector field as:
\[
\resizebox{.95\linewidth}{!}{$
F = \begin{bmatrix}
    x_2[1] & x_2[2] & \cdots & x_2[N] \\
    -x_1[1]+x_1[1]^3 & -x_1[2]+x_1[2]^3 & \cdots & -x_1[N]+x_1[N]^3
\end{bmatrix}.
$}
\]
The corresponding Gram matrix becomes
\[
\resizebox{.95\linewidth}{!}{$
G = FF^{\top} = \begin{bmatrix}
    \sum\limits_{i=1}^{N}x_2[i]^2 & -\sum\limits_{i=1}^{N}x_2[i](x_1[i] - x_1[i]^3) \\
    -\sum\limits_{i=1}^{N}x_2[i](x_1[i] - x_1[i]^3) & \sum\limits_{i=1}^{N}(x_1[i] - x_1[i]^3)^2
\end{bmatrix}.
$}
\]
Mechanically, \( G_{11}=\sum_{i=1}^{N}x_2[i]^2 \) is proportional to the kinetic energy of the system at the grid points, while \( G_{22}=\sum_{i=1}^{N}\left(x_1[i]^2 -2x_1[i]^4+x_1[i]^6\right)\approx \sum_{i=1}^{N}x_1[i]^2 \) approximates the potential energy in the vicinity of \( \mathbf{x}=\mathbf{0} \). Consequently, the trace \(\mathrm{tr}(G)\) provides a local approximation of the total energy near the origin, demonstrating that \(F\) and \(G\) capture two distinct types of dynamical features in this example. It should be emphasized that for non-Hamiltonian systems, the style feature defined above does not necessarily encode an energy.
 \subsection{\codexnew{Transformation Properties of Style Features}}
\label{sec:style}
\codexnew{The distinction between content and style can already be seen at the input layer. Under time reversal, the vector field tensor changes from \(V\) to \(-V\), so its pointwise content changes, whereas the Gram representation is exactly preserved because \((-V)(-V)^\top=VV^\top\). A coordinate relabeling similarly permutes the spatial samples and vector field channels; after the corresponding channel alignment, the Gram representation is unchanged even though the pointwise tensor is rearranged.}

\codexnew{Reflections and rotational conjugacies are not exact invariances of the Gram representation in general. Nevertheless, across the Lorenz, Chen, Lv, and R\"ossler examples considered here, their Gram discrepancies remain smaller than the associated pointwise discrepancies. These observations establish a limited and physically interpretable transformation stability rather than universal invariance. The definitions, transformation protocol, and complete numerical results are provided in Appendix~\ref{app:style_robustness}.}
 \subsection{\codexnew{Fusion of Lorenz-like Systems Using Content and Style Features}}
\label{sec:lorenz_features}
\begin{figure}[tb]
\centering
\includegraphics[width=0.48\textwidth]{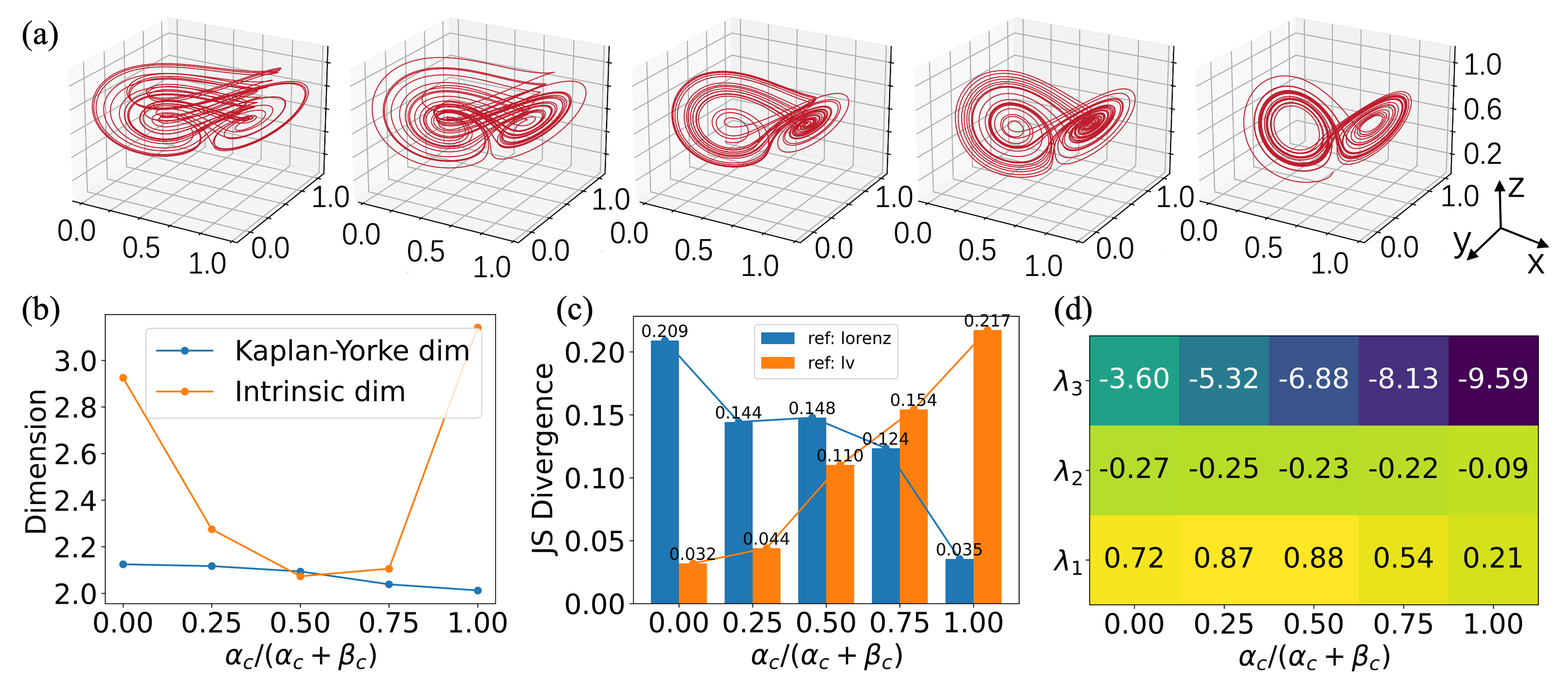}
\caption{Fusion of the Lorenz and Lv's systems for varying content weights. (a) Phase space trajectories of the fused systems. From left to right, the fusion parameter $\alpha_c/(\alpha_c+\beta_c)$ is 0.00, 0.25, 0.50, 0.75, and 1.00. (b)-(d) Dynamical signatures trace out landscapes as the fusion parameter varies.
Here, each system is normalized in time and state space; refer to Appendix~\ref{app:architecture} for detailed configurations.}
\label{lorenz+lv}
\end{figure}

We consider three canonical chaotic systems, the Lorenz, Chen's, and Lv's systems, which form a generalized Lorenz family with vector fields containing at most second-order terms~\cite{lorenz_like2005csf}. For brevity, we refer to them as Lorenz-like systems.
First, we explore the fusion of the Lorenz and Lv's systems. To assess whether the content features faithfully capture system structure, we perform fusion using content features only. We obtain explicit governing equations by linear regression on the fused vector fields and then perform numerical simulations. As shown in Fig.~\ref{lorenz+lv}\hyperref[lorenz+lv]{(a)}, increasing the content weight assigned to one parent system drives the descendant attractor toward that parent’s attractor. We further compute various dynamical signatures of the fused system, including its intrinsic and Kaplan-Yorke dimensions, the Jensen-Shannon divergence between its invariant measure distribution and that of the parent system, and its Lyapunov spectrum. As shown in Figs.~\ref{lorenz+lv}\hyperref[lorenz+lv]{(b)–(d)}, these dynamical signatures vary smoothly with the fusion parameter, indicating that the content features largely preserve the dynamics of the parent systems.

To further investigate the role of style features in system fusion, we fuse the content of the Lorenz system with the style of Chen's system while varying the mixing ratio \(\alpha_c/(\alpha_c+\beta_s)\). During this process, the Kaplan-Yorke dimension as well as the second and third Lyapunov exponents of the fused system remain nearly invariant.  Conversely, the maximal Lyapunov exponents (MLEs) exhibit dramatic variation. As summarized in Tab.~\ref{table1}, the fused system first generates chaotic dynamics and subsequently undergoes a transition of gradual loss of chaos as the contribution of Chen's style increases.

Taken together, these fusion experiments on representative models suggest that {\it content shapes the attractor geometry, whereas style controls the statistical correlation structure of the vector field across both state components and spatial locations}, thereby inducing or suppressing chaos as measured by the MLE.

\begin{table}[b]
  \caption{\label{table1}MLEs of the fused system under varying proportions of content (Lorenz) to style (Chen's) features.}
  \centering
  \footnotesize
  \renewcommand{\arraystretch}{1.25}
  \begin{ruledtabular}
    \begin{tabular}{c|ccccccccc}
      $\frac{\alpha_c}{\alpha_c+\beta_s}$ & 0.00 & 0.01 & 0.10 & 0.25 & 0.50 & 0.75 & 0.90 & 0.99 & 1.00 \\
      \hline
      MLE & -1.054 & 0.007 & 0.220 & 0.234 & 0.271 & 0.247 & 0.247 & 0.305 & 0.288 \\
    \end{tabular}
  \end{ruledtabular}
\end{table}

 \subsection{Layerwise perturbation analysis}\label{sec:perturbation}
To probe the physical meaning of features at different layers, we perform perturbations of the features and quantify the resulting changes in dynamical signatures of the reconstructed system. Specifically, we trained a three-dimensional CAE on discretized vector fields of the Lorenz system sampled on a fixed uniform grid. After training, given an input vector field, we perturb the pre-activation tensor at the chosen layer by adding zero-mean Gaussian noise. The noise amplitude is set to a prescribed fraction $\epsilon$ of the root mean square (RMS) value of that tensor computed from the unperturbed forward pass. We then propagate the perturbed activation through the remaining layers of the encoder and the decoder to obtain a reconstructed vector field. For each layer and each $\epsilon$, we repeat this procedure over multiple random seeds and record the reconstructions.

To obtain a physical interpretation, we projected the reconstructed field onto the Lorenz family by linear regression of the parameters $(\sigma,\rho,\beta)$ in the standard Lorenz form $\dot x=\sigma(y-x)$, $\dot y=\rho x-xz-y$, and $\dot z=xy-\beta z$ over the full grid. We report the normalized regression residual as a measure of consistency with the skeleton of the Lorenz system, together with the mean-squared reconstruction error relative to the input field. From the regressed parameters we compute the dissipativity proxy $D=\sigma+1+\beta$ and its change relative to the unperturbed reconstruction. We also compute the Lyapunov exponents, the Kaplan Yorke dimension for each regressed system.

\begin{figure}[htb]
\centering
\includegraphics[width=0.48\textwidth]{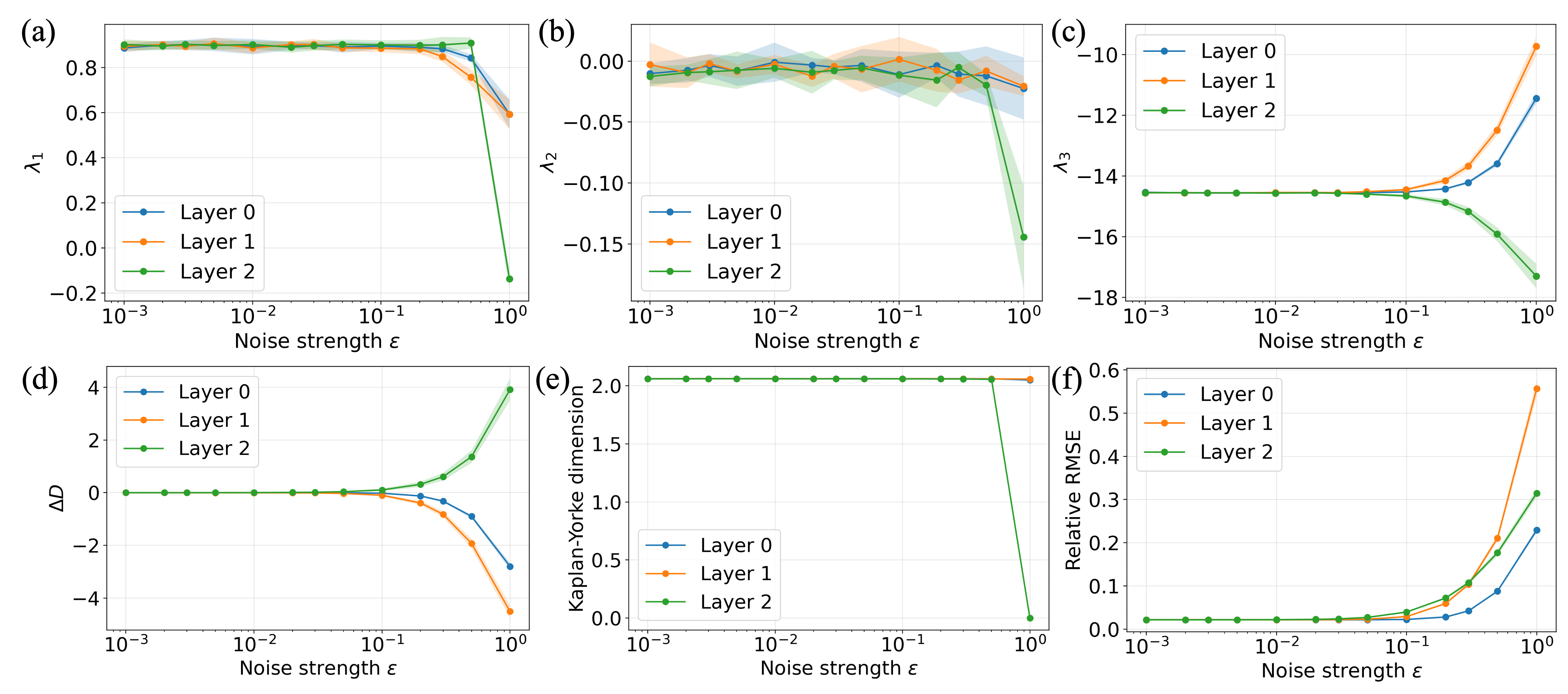}
\caption{Changes in the dynamical signatures of the reconstructed Lorenz system after adding Gaussian noise to features at different layers. (a)–(c) Lyapunov exponents, with $\lambda_1>\lambda_2>\lambda_3$. (d) Change in dissipativity, $\Delta D$, relative to the classical Lorenz system, where $D = -\nabla \cdot f$. (e) Kaplan-Yorke dimension. (f) Normalized residual of the linear regression relative to the discretized vector field.}
\label{responds_exp}
\end{figure}

Figure~\ref{responds_exp} summarizes the results and shows that features at different layers control different dynamical signatures and thus carry different physical meanings. Different layers influence these signatures with different strengths and even in opposite directions. In particular, The zeroth layer has the weakest effect, consistent with a low-level local representation; the first layer weakens dissipativity (Fig.~\ref{responds_exp}\hyperref[responds_exp]{(c-d)}) and significantly influences the skeleton of the Lorenz attractor (Fig.~\ref{responds_exp}\hyperref[responds_exp]{(f)}); and the second layer enhances dissipativity (Fig.~\ref{responds_exp}\hyperref[responds_exp]{(c-d)}) and, under sufficiently strong perturbations, determines whether chaos is maintained (Fig.~\ref{responds_exp}\hyperref[responds_exp]{(a)}). In addition, over a substantial noise strength, the Kaplan–Yorke dimension remains nearly unchanged (Fig.~\ref{responds_exp}\hyperref[responds_exp]{(e)}), suggesting that the CAE preserves information about the attractor dimension within that regime.

 \FloatBarrier

\section{\codexnew{NDTL-induced Classification}}
\label{sec:organization}

\subsection{\codexnew{Dynamical Signature Landscapes and System Dissimilarity}}
\label{sec:classification}
\begin{figure}[tb]
\centering
\includegraphics[width=0.48\textwidth]{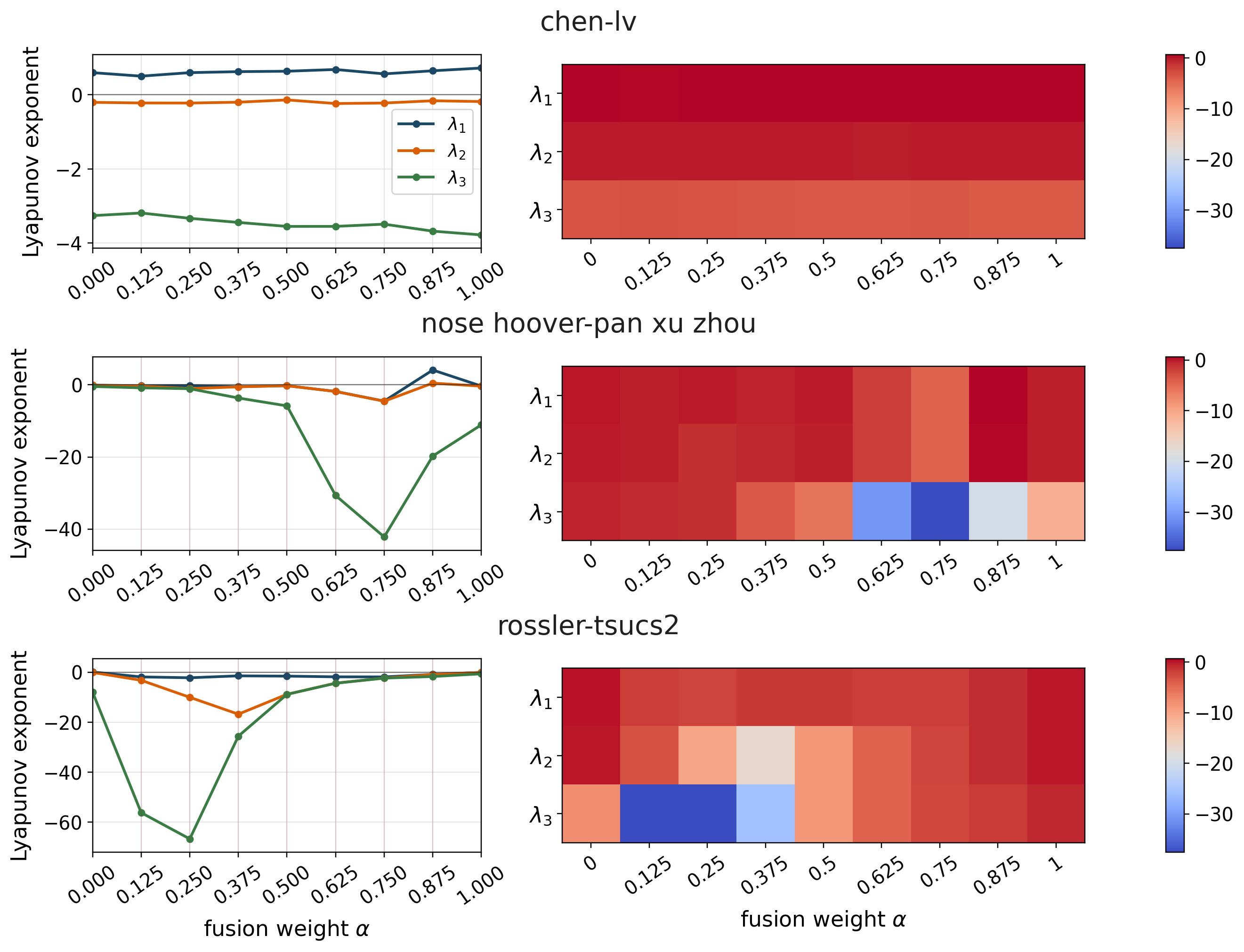}
\caption{\codexnew{Representative small distance and large distance Lyapunov spectrum landscapes. The small distance pair exhibits a smooth spectrum landscape and preserves nontrivial chaotic behavior over most fusion weights. The large distance pair shows a distorted spectrum landscape and loses chaotic behavior over a larger portion of the fusion interval. These examples illustrate how the NDTL-induced distance reflects the continuity of the generated dynamics.}}
\label{signature_landscape}
\end{figure}

As the relative weights of the parent systems change, the generated descendants trace out landscapes of dynamical signatures, reflecting the closeness between the parent systems. Smooth landscapes indicate that the two systems are close in this sense, whereas collapsed attractors, loss of chaos, or strongly distorted landscapes \codexnew{(Fig.~\ref{signature_landscape})} indicate larger separation. Thus, NDTL naturally induces a principled criterion for quantifying dynamics similarity and, in turn, enables dynamics classification.

To make this criterion quantitative, we use the Lyapunov spectrum as the dynamical signature in the classification experiment. For each pair of parent systems, content feature fusion is performed over a sequence of fusion weights, producing a family of descendant vector fields. The resulting Lyapunov spectra form a spectrum landscape.
 For each unordered pair \((A,B)\), we train a CAE using only the two endpoint vector fields. The fusion is performed using content features only. Let \(X_A\) and \(X_B\) be the normalized vector field tensors of the two systems, and let \(F_k(X)\) denote the encoder feature at layer \(k\). The content losses are
\begin{equation*}
\begin{aligned}
    L_A&=\frac{1}{3}\sum_{k=1}^{3}\|F_k(X)-F_k(X_A)\|_2^2,\\
    L_B&=\frac{1}{3}\sum_{k=1}^{3}\|F_k(X)-F_k(X_B)\|_2^2 .
\end{aligned}
\end{equation*}

For a fusion parameter \(\alpha\), the optimized field minimizes
\begin{equation*}
L=(1-\alpha)L_A+\alpha L_B .
\end{equation*}

We use nine values of \(\alpha\), namely
\begin{equation*}
\alpha\in\{0,0.125,0.25,0.375,0.5,0.625,0.75,0.875,1\}.
\end{equation*}
The endpoint fields at \(\alpha=0\) and \(\alpha=1\) are not optimized, while the seven intermediate fields are initialized by linear interpolation and then optimized by content feature matching. For every optimized fusion path, we compute the Lyapunov spectrum
\begin{equation*}
\lambda(\alpha)=\big(\lambda_1(\alpha),\lambda_2(\alpha),\lambda_3(\alpha)\big).
\end{equation*}

We define the pairwise distance from four properties of the resulting Lyapunov spectrum path. The endpoint gap is
\begin{equation*}
D_{\rm end}=\|\lambda(1)-\lambda(0)\|_2 .
\end{equation*}
The barrier term measures deviation from the linear interpolation between endpoint spectra,
\begin{equation*}
D_{\rm bar}=\left\langle
\left\|
\lambda(\alpha)-\big[(1-\alpha)\lambda(0)+\alpha\lambda(1)\big]
\right\|_2
\right\rangle_{0<\alpha<1}.
\end{equation*}
The roughness term measures the second finite difference of the spectrum path,
\begin{equation*}
D_{\rm rough}=
\left\langle
\|\lambda_{i+1}-2\lambda_i+\lambda_{i-1}\|_2
\right\rangle_i .
\end{equation*}
Finally, the chaos loss term is the fraction of fusion weights for which the Lyapunov computation fails or the largest Lyapunov exponent is nonpositive.

Each component is converted to a percentile rank over all evaluated pairs. The final NDTL-induced distance is
\begin{equation*}
\begin{aligned}
d(A,B)=\frac{1}{4}\big[&
r(D_{\rm end})+r(D_{\rm bar})\\
&+r(D_{\rm rough})+r(D_{\rm chaos})\big].
\end{aligned}
\end{equation*}
Small values of \(d(A,B)\) therefore indicate that two systems have similar endpoint spectra, a smooth spectrum path, and few losses of chaotic behavior during fusion.

Importantly, the induced similarity depends on the chosen feature constraints. In the benchmark experiment, we use content features to define dynamical proximity based on content feature fusion. Alternative choices of feature terms in the objective can induce different similarity relations.

\subsection{\codexnew{Classification of Chaotic Systems}}
\label{sec:benchmark_classification}
The benchmark systems are selected from the chaotic systems database of Ref.~\cite{gilpin_benchmark_nips2021}. The original benchmark contains 131 systems. After excluding time-delay systems and systems incompatible with the present vector field representation, we retain 52 three-dimensional continuous time chaotic systems. The selected systems and scripts for reproducing the analysis are provided in the code repository described in the Data and Code Availability statement. For each system, an attractor box is estimated from a representative trajectory. The box is then affinely mapped to \([0,1]^3\), and the vector field is scaled so that the mean vector norm on the \(48^3\) grid equals \(5.0\). Therefore, the classification analysis compares normalized vector fields rather than raw equations in their original coordinates.

\codexnew{Applying the construction above to every pair produces a symmetric NDTL-induced dissimilarity matrix. Denoting its entries by $D_{ij}$, we convert this matrix into an affinity matrix suitable for spectral clustering:}
\begin{equation*}
W_{ij}=\exp\left(-\frac{D_{ij}^2}{2\sigma^2}\right),
\end{equation*}
where \(\sigma\) is the median of the nonzero pairwise distances between distinct systems. Clustering labels are obtained from this affinity matrix. The two-dimensional spectral embedding used for visualization is also computed from the same affinity matrix, but the clustering labels are not determined by the two-dimensional display.

\begin{figure}[!b]
\centering
\includegraphics[width=\columnwidth]{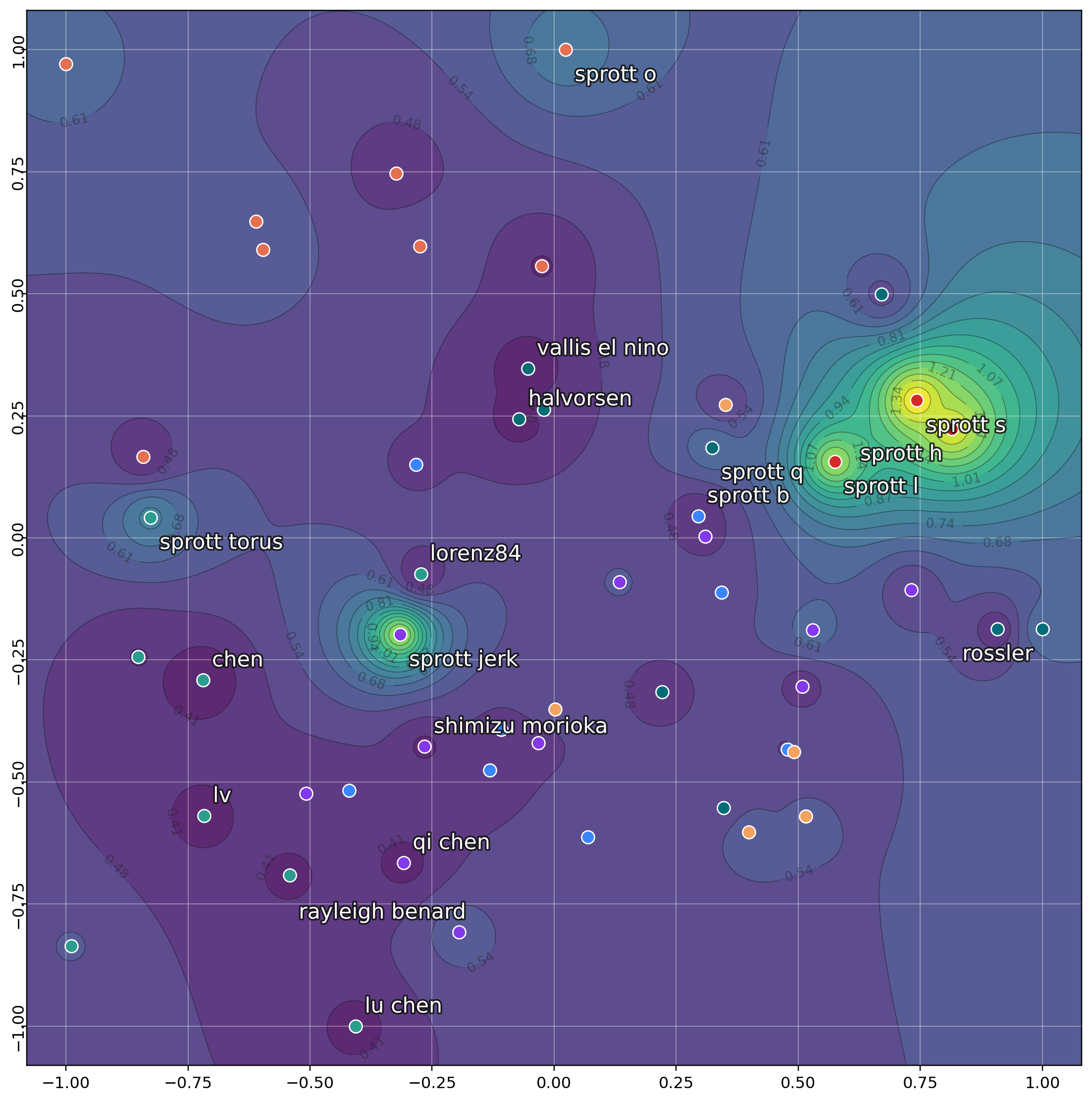}
\caption{\codexnew{Two-dimensional spectral embedding of the NDTL-induced affinity matrix for the chaotic system benchmark. Each point represents one system, and colors denote the seven clusters. Nearby points correspond to systems with lower NDTL-induced dissimilarities, whereas isolated points have larger average dissimilarities from the rest of the benchmark.}}
\label{fig:classification_results}
\end{figure}

The embedding in Fig.~\ref{fig:classification_results} provides an exploratory classification of the benchmark based on the behavior of generated descendants. The Lorenz, Chen, and Lv systems occupy nearby regions, with Chen and Lv forming a close pair, which is consistent with the smooth Lyapunov spectrum paths observed within the Lorenz-like family. In contrast, the R\"ossler system, the Nos\'e-Hoover system, and several Sprott systems lie in relatively sparse or boundary regions, indicating stronger spectral distortion or more frequent loss of chaos during fusion with other systems. Accordingly, systems that are close in the embedding tend to exhibit more continuous Lyapunov spectrum changes under the present fusion protocol, whereas isolated systems more often yield distorted spectra or lose chaos.

\subsection{\codexnew{Comparison of Spectral and Hierarchical Clustering}}
\label{sec:clustering_consistency}
\codexnew{The NDTL-induced dissimilarity matrix is the primary object in this classification, whereas spectral clustering is one downstream partition of that matrix. We therefore compare the spectral result with \codexflag{average-linkage} hierarchical clustering to examine which relations persist under a different clustering scheme. For two clusters \(C_a\) and \(C_b\), the hierarchical distance is}
\begin{equation*}
d(C_a,C_b)=
\frac{1}{|C_a||C_b|}
\sum_{i\in C_a}\sum_{j\in C_b}d(i,j).
\end{equation*}

\begin{figure*}[t]
\centering
\includegraphics[width=\textwidth]{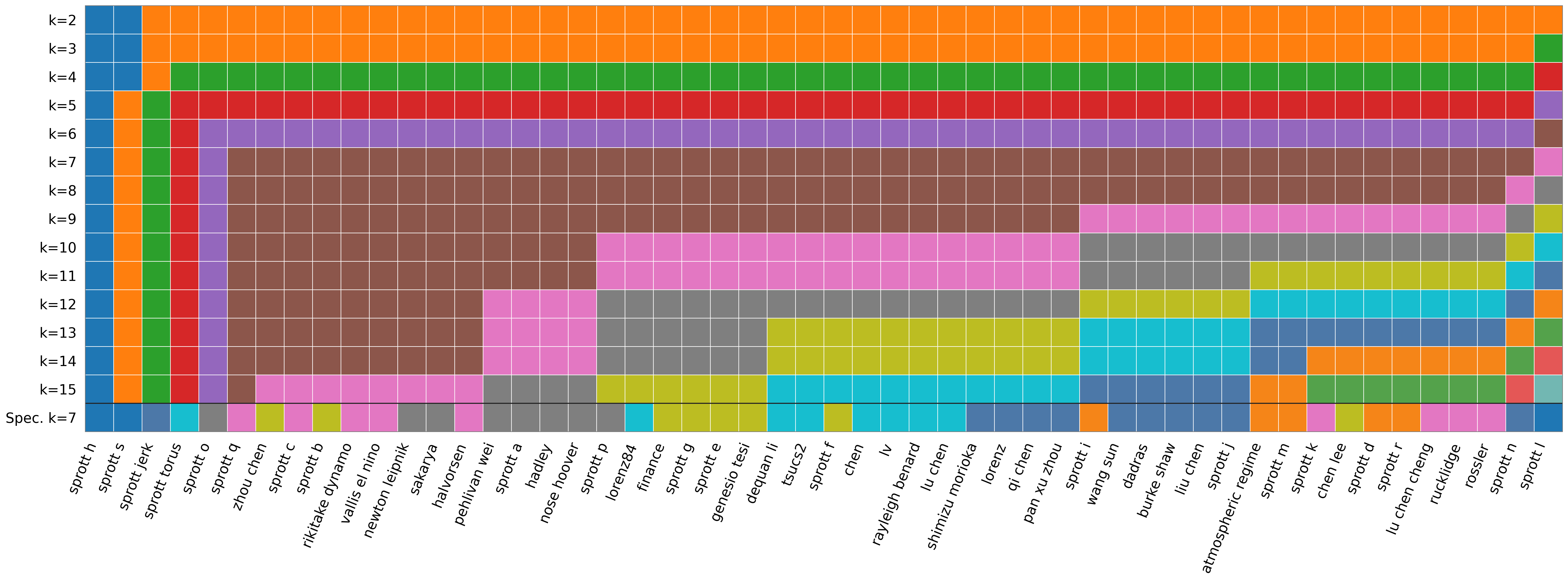}
\caption{Comparison between hierarchical clustering and spectral clustering for the retained systems from the benchmark of chaotic systems. Each row \(k=2,\ldots,15\) shows the \codexflag{average-linkage} hierarchical clustering assignment for a chosen number of clusters, while the bottom row shows the spectral clustering assignment with \(k=7\) on the same system ordering.}
\label{fig:classification52_hierarchical}
\end{figure*}

Figure~\ref{fig:classification52_hierarchical} compares \codexflag{average-linkage} hierarchical clustering with the spectral clustering assignment. Both methods use the same distance matrix. Hierarchical clustering preserves the nested organization of the distance structure, whereas spectral clustering emphasizes group structure in the affinity graph. Their common groupings support the consistency of the induced classification, while their differences reflect how distinct clustering methods partition a continuous distance structure.

\codexnew{The persistence of several groupings across cluster numbers and clustering schemes indicates that the organization is carried by the NDTL-induced dissimilarities rather than by a particular two-dimensional display. The resulting classification therefore reflects the behavior of dynamics generated along fusion paths between vector fields.}

\FloatBarrier
 \FloatBarrier

\section{\codexnew{Applications of NDTL}}
\label{sec:applications}
\codexnew{We next demonstrate applications of systems created by NDTL across multiple domains.}
\subsection{\codexnew{Application to the Hastings-Powell food chain model}}
\textcolor{orange}{We use the Hastings-Powell (HP) food chain model to test whether a vector field generated by NDTL can be projected back into an interpretable ecological model. The HP model describes a resource \(P\), a herbivore \(H\), and a carnivore \(C\)~\cite{chaotic_rcp1991},}
\[
\color{orange}
\begin{aligned}
\dot P &= P\left(1-\frac{P}{K}\right)-u_h(P)H,\\
\dot H &= H\left(u_h(P)-d_h\right)-u_c(H)C,\\
\dot C &= C\left(u_c(H)-d_c\right),
\end{aligned}
\]
\textcolor{orange}{where the saturating uptake functions are}
\[
\color{orange}
u_h(P)=\frac{\alpha_h P}{\gamma_h+P},\quad
u_c(H)=\frac{\alpha_c H}{\gamma_c+H}.
\]
\textcolor{orange}{Previous work on the HP model has shown that, for some specific parameter values, the model can exhibit stable, periodic, and chaotic coexistence~\cite{chaotic_rcp1991}.}

\begin{figure}[tb]
\centering
\includegraphics[width=\linewidth]{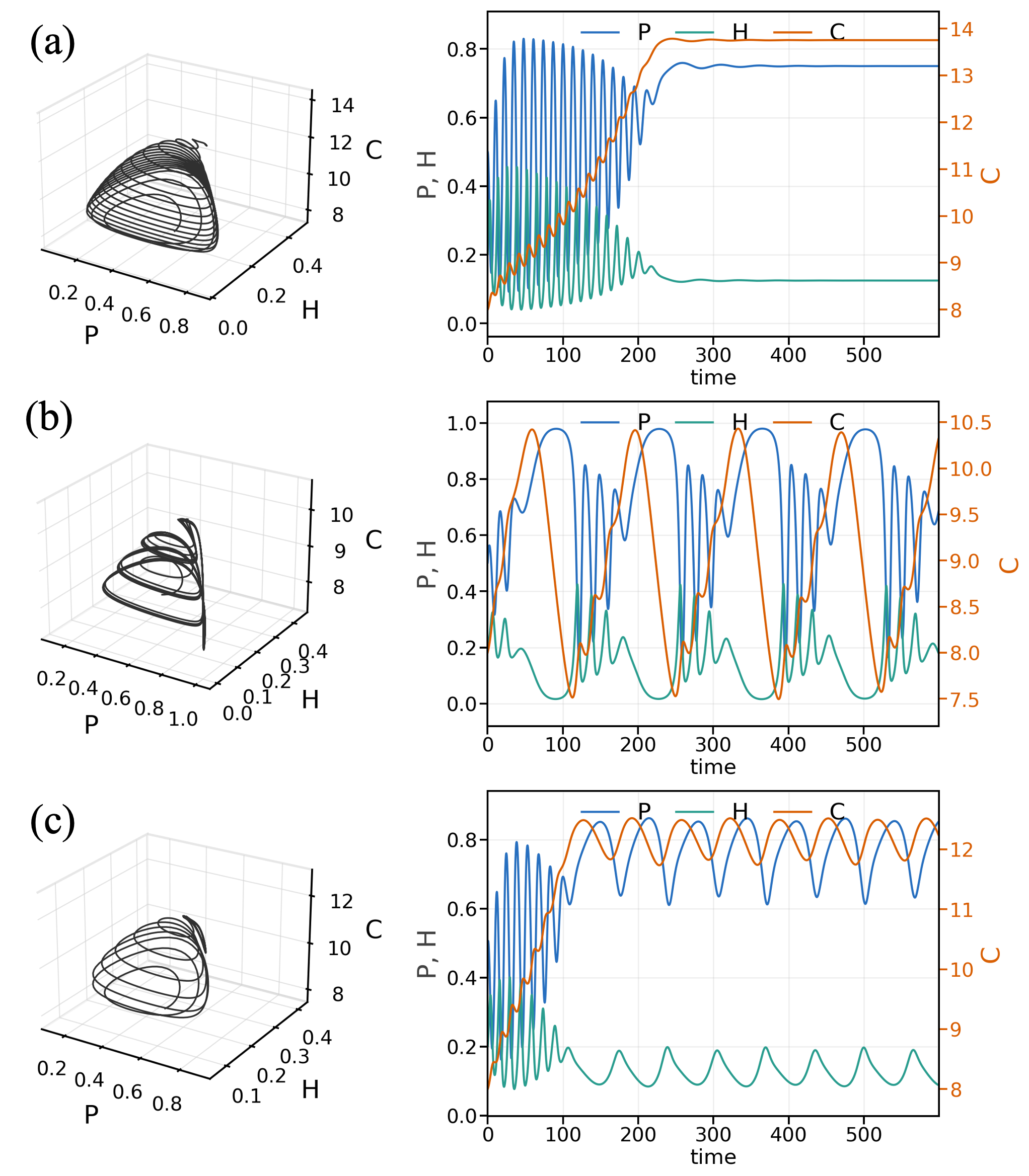}
\caption{\textcolor{orange}{
Ecological system creation in the Hastings-Powell food chain model. \textcolor{orange}{(a) The first parent system converges to a coexistence equilibrium. (b) The second parent system exhibits chaotic coexistence of the three species.} (c) The system generated by NDTL at \(\alpha_b=0.50\) approaches a limit cycle after a transient phase. Thus, NDTL creates persistent oscillatory coexistence among three species from stable and chaotic ecological parents.
}}
\label{hp}
\end{figure}

\textcolor{orange}{We select two systems with distinct dynamical properties as parent systems. The first system, with \((\alpha_h,\gamma_h)=(2.500,0.500)\), converges to a coexistence equilibrium, whereas in the second system, with \((\alpha_h,\gamma_h)=(1.667,0.333)\), the three species persist on a chaotic attractor, as shown in Fig.~\ref{hp}\hyperref[hp]{(a)} and Fig.~\ref{hp}\hyperref[hp]{(b)}, respectively.}

\textcolor{orange}{We fuse the two systems using their content features and use symbolic regression to project the resulting vector field onto the HP model family. At the fusion weight of \(\alpha_b=0.50\), the projected system has \(\alpha_h=2.0766\) and \(\gamma_h=0.4234\). As shown in Fig.~\ref{hp}\hyperref[hp]{(c)}, the fused system retains a similar ecological interpretation to the parent systems: the abundances of the three species are comparable to those of the parents, and all three trophic levels coexist in the long term. However, unlike either the coexistence equilibrium of the first parent or the chaotic attractor of the second parent, after a transient period the three populations in the fused system converge to a limit cycle. This experiment shows that, starting from two ecological systems with different long-term behaviors, NDTL can generate through fusion an interpretable oscillatory coexistence regime, rather than obtaining it through a mechanical parameter scan. The training protocol, the projection procedure, validation over multiple initial conditions, and Floquet analysis are given in Appendix~\ref{app:ecology}.}

 \subsection{\codexnew{Application to the SEIRS epidemiological model}}
\label{sec:extra_domain_fusion}
Here we report an additional example involving an epidemiological model. This example uses the same feature matching, vector field optimization, and regression pipeline, but is not used in the classification benchmark.

We further explore the applicability of the NDTL to integrate the Lorenz system with the Susceptible Exposed Infectious Recovered Susceptible (SEIRS) system, an epidemiological model~\cite{seir2015}.
Here, the SEIRS system is described by the equations:
\begin{equation}
\label{seir_simple}
\begin{array}{l}
\dot{S}  = -\beta \frac{SI}{N} + \delta R, \quad \dot{E} = \beta \frac{SI}{N} - \sigma E, \\ 
\dot{I}  = \sigma E - \gamma I, \quad \dot{R} = \gamma I - \delta R,
\end{array}
\end{equation}
where \(S\), \(E\), \(I\), and \(R\) represent the susceptible, exposed, infectious, and recovered population sizes, respectively, and \(N = S + E + I + R\) is the total population. The parameters \(\beta\), \(\sigma\), \(\gamma\), and \(\delta\) denote the transmission rate, incubation rate, recovery rate, and immunity waning rate, respectively.

We interpret the variables in the SEIRS system as fractions of the total population, subject to the constraint \(S + E + I + R = N = 1\), thereby reducing the SEIRS system to an effective three-dimensional system. We fuse this SEIRS system by matching its content feature with the style feature of the Lorenz system. During optimization, we use the discretized vector field corresponding to the first three dimensions of the SEIRS model. In the regression procedure, to achieve a sparser and more interpretable set of equations, we employ the LASSO regression.
However, due to the implicit dependence on the variable \(R\), the exact contribution of \(R\) within the SEI equations cannot be directly inferred. Drawing on prior epidemiological knowledge, specifically, that the exposed (\(E\)) and infected (\(I\)) compartments generally do not arise directly from the recovered compartment (\(R\)), we impose the assumption that the dynamics governing \(E\) and \(I\) are independent of \(R\). With this assumption in place, we fix the style weight associated with the Lorenz system and vary the content weight of the SEIRS system, thereby obtaining interpretable epidemiological models across a broad parameter range.

Specifically, the fused equations of the SEIRS and Lorenz system are obtained via the LASSO regression as:
\begin{equation}
\label{seir_verbose}
\begin{array}{l}
\dot{S}  = f(S, E, I, R), \quad \dot{E} = \kappa S -\sigma_1 E + \beta \frac{SI}{N}, \\ 
\dot{I}  = \sigma_2 E - \gamma I + \xi \frac{E^2}{N}, \quad \dot{R} = -\dot{S} - \dot{E} - \dot{I}.
\end{array}
\end{equation}
where the parameters \(\beta\) and \(\gamma\) retain their original epidemiological interpretations from the SEIRS system~\eqref{seir_simple}. Compared to the classical equations, this model introduces an additional nonlinear infection term \(\xi \frac{E^2}{N}\) in the infected compartment equation, capturing potential secondary infection or superinfection effects at high exposure densities. The strength of this nonlinear interaction is quantified by the parameter \(\xi\). Additionally, the exposed compartment equation now includes an extra linear source term \(\kappa S\), representing an influx into the exposed class independent of direct infectious contact. These modifications reflect more complex epidemic dynamics than those accounted for in the classical formulation.

\begin{figure}[t]
  \centering
  \includegraphics[width=0.48\textwidth]{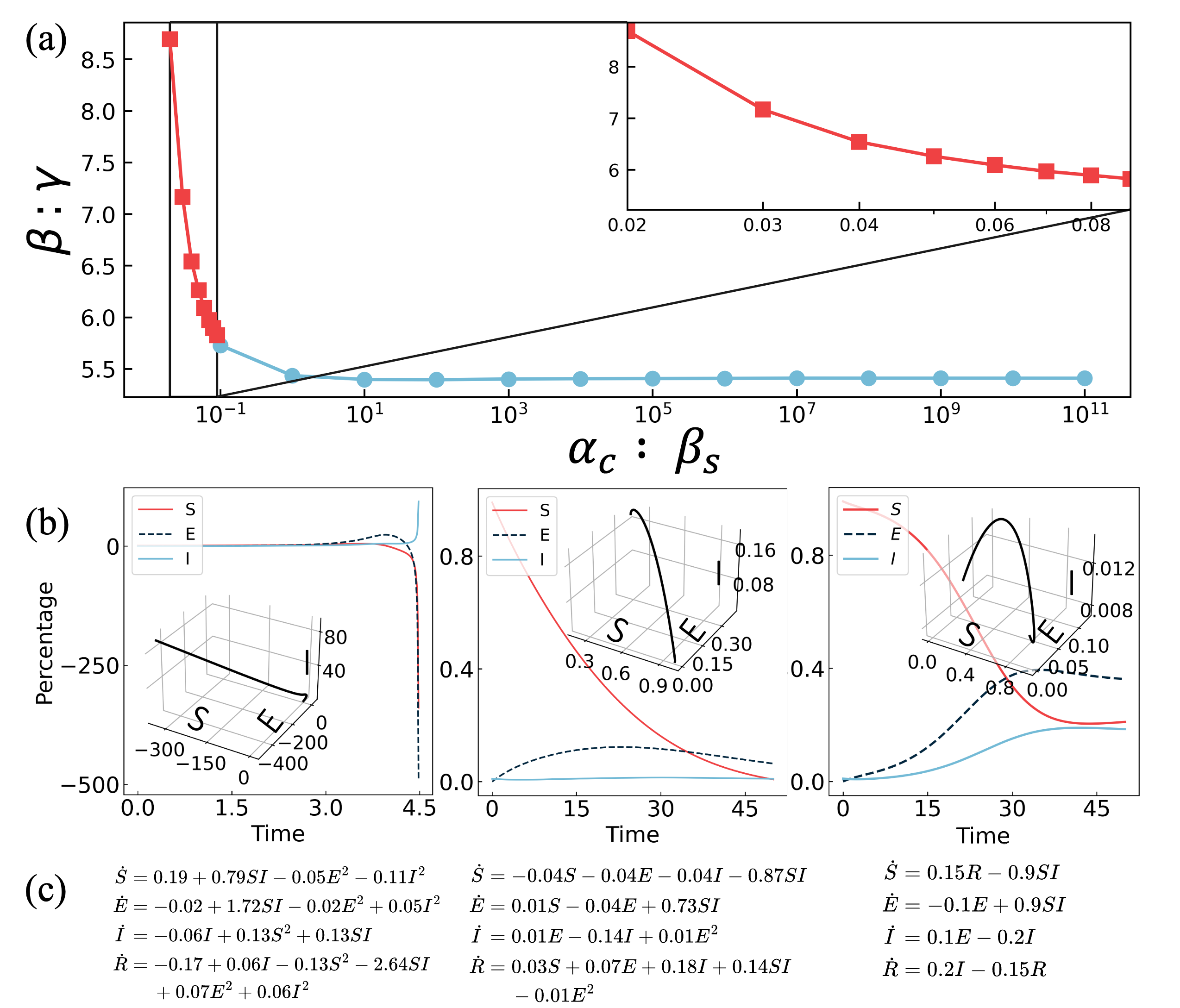}
  \caption{Fusion of the Lorenz system and SEIRS system under varying optimization weights. (a) 
  Index values $\frac{\beta}{\gamma}$ across different feature weight ratios. (b) The trajectories of the newly created SEIRS system at three representative fusion stages (from left to right: \( \frac{\alpha_c}{\beta_s} \leq 10^{-2}\), =\(10^{-5}\), and the initially trained SEIRS system). (c) Governing equations obtained via the LASSO regression.}
  \label{seris}
\end{figure} 

A fundamental metric in infectious disease epidemiology is the basic reproduction number \( R_0 \), defined as the average number of secondary infections produced by a single infected individual in a fully susceptible population, in the absence of interventions~\cite{math_disease2008nrm}. For the classical system~\eqref{seir_simple}, this quantity is directly given by \( R_0 = \frac{\beta}{\gamma} \). In contrast, for the created SEIRS system~\eqref{seir_verbose}, although the ratio \(\frac{\beta}{\gamma}\) no longer precisely equals the overall basic reproduction number due to additional flow terms and nonlinear infection effects, it still retains an important epidemiological interpretation. Specifically, the ratio \(\frac{\beta}{\gamma}\) characterizes the intrinsic balance between transmission and recovery rates, representing the pathogen's inherent potential to spread under idealized conditions, namely, when secondary infection effects and transitions involving exposed states are negligible.

In Fig.~\ref{seris}, we show the representative results of the system fusion experiments with the SEI initial conditions \((0.99,\,0,\,0.01)\). For \(\frac{\alpha_c}{\beta_s} \leq 10^{-2}\), the created system loses the fundamental characteristics of the original SEIRS model due to excessive interference from the Lorenz style features. Meanwhile, the absence of Lorenz content features prevents the system from exhibiting chaotic dynamics. For ratios \(\frac{\alpha_c}{\beta_s}\) within the range \(10^{-2}\) to \(10^{11}\), the regression on the discretized vector fields yields equations consistent with the equations in \eqref{seir_verbose}.
Moreover, as the proportion of Lorenz style features increases, the fused system gradually approaches scenarios characterized by higher values of \(\frac{\beta}{\gamma}\).
 \subsection{\codexnew{Application to chaotic image encryption}}
\textcolor{orange}{Finally, we test whether a created chaotic system can be used as a source in a standard image encryption procedure. Chaotic trajectories are often used in permutation diffusion encryption because they can generate deterministic but sensitive key streams. We use the fused system generated in Sec.~\ref{sec:lorenz_features} by combining the content of the Lorenz system with the style of Chen's system as the chaotic source. Starting from a fixed initial condition, its trajectory is converted into a permutation sequence and an 8 bit diffusion mask. The image is first permuted and then diffused twice; decryption applies the inverse operations with the same key.}

\begin{figure}[t]
    \centering
    \includegraphics[width=0.98\linewidth]{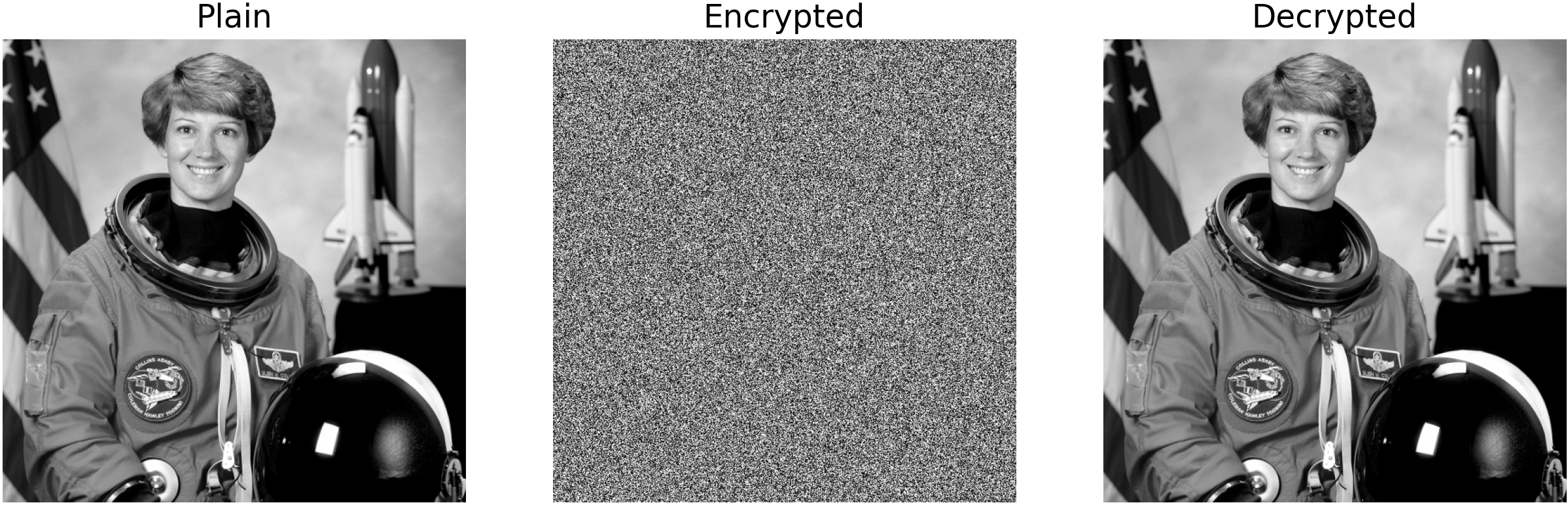}
    \caption{\textcolor{orange}{Encrypted and decrypted images for the fused system generated by combining the content of the Lorenz system with the style of Chen's system. The permutation diffusion procedure produces visually random ciphertext and reconstructs the original image with the correct key.}}
    \label{fig:crypto_fused}
\end{figure}

\textcolor{orange}{Figure~\ref{fig:crypto_fused} shows the result for this fused system. The encrypted image is visually indistinguishable from random noise, while the correct key reconstructs the original image exactly. Thus, the generated vector field is not only a dynamical object for analysis, but also a source that can drive a complete downstream algorithm.}

\textcolor{orange}{The quantitative indicators are also consistent with a valid chaotic encryption source. The fused system gives cipher entropy \(7.999325\), adjacent pixel correlation \(0.001306\), NPCR \(99.613\%\), UACI \(33.448\%\), key NPCR \(99.614\%\), and an aggregate score \(0.999484\). These values are close to the standard targets for random looking ciphertext and plaintext or key sensitivity. Candidate selection, comparison with the parent systems, and full image level metrics are given in Appendix~\ref{app:encryption}.}

 \FloatBarrier

\section{\codexnew{Concluding Remarks}}
\label{sec:discussion}
In this Article, we introduce NDTL, a framework at the interface of machine learning and dynamical systems theory for creating dynamical systems that combine prescribed behaviors from multiple parents. \textcolor{orange}{In the Lorenz-like examples, the generated systems exhibit systematic changes in several dynamical signatures, and the two feature representations have empirically distinct effects within the present construction. We further used Lyapunov spectrum trajectories along content fusion paths to define a representation based on fusion path dissimilarity and to obtain an unsupervised organization of a subset of chaotic systems. Finally, applications to ecological and epidemiological modeling and chaotic image encryption demonstrate the broader applicability of NDTL.}

In addition, the layerwise perturbation experiments in Sec.~\ref{sec:perturbation} indicate that features at different layers influence dynamical signatures with different magnitudes and, in some cases, opposite effects, pointing to layer-wise physical interpretations. Leveraging this controllability for physics-informed system design and modulation is a promising direction for future work.

Finally, from a computational perspective, the CAE maps discretized vector fields to a compact latent space, which mitigates the high ambient dimensionality. When one is willing to assume a fixed parametric form a priori, directly optimizing sparse equation parameters remains a viable and potentially more efficient alternative.
 
\clearpage
\section*{\codexnew{Data and Code Availability}}
All the implementations described above are publicly available and fully reproducible. The complete source code can be accessed through the GitHub repository associated with this study: \url{https://github.com/967255/NDTL}.

\appendix
\section{\codexnew{NDTL Architecture, Training, and Preprocessing}}
\label{app:architecture}

To extract the low-dimensional structure from the three-dimensional (3D) vector field, we employ a convolutional autoencoder (CAE), which comprises  
\(\mathcal{I}\), a nonlinear encoder that compresses the input into a compact latent representation, and  
\(\mathcal{D}\), a symmetric decoder that reconstructs the original field from that representation.

\paragraph{Vector Field Preprocessing and Hyperparameter Configuration}  
As an illustrative example of the 3D system, let \(X:[a,b]^3\to\mathbb{R}^3\) be a continuous vector field, sampled on a uniform grid \(D\times H\times W\) to produce $X \in \mathbb{R}^{3\times D\times H\times W},$ which serves as the CAE input. We set $D=H=W=48$ so that each down-sampling stride divides the grid exactly; empirical tests confirm that this resolution delivers adequate reconstruction quality.

The CAE encoder \(\mathcal I\) consists of \(L=3\) strided 3D convolutions with channel widths $c_0=3$ and $c_i=2^i\times64$ for $i=1,2,3$. We set $(k_i,s_i,p_i)=(5,3,2),\;(3,2,1),\;(3,2,1)$
for the kernel size, stride, and padding at layers \(i=1,2,3\). The decoder \(\mathcal D\) mirrors this structure using transposed convolutions with the same \(\{k_i,s_i,p_i\}\).

\paragraph{Encoder}  
Let \(Z^{(0)}=X\), and denote by  
$n_i=\frac{48}{\prod_{j=1}^i s_j}$ the spatial resolution after layer \(i\) for $i=1,2,3$.  
Then each encoder layer \(\mathcal I_i\) perform as:
\[
\begin{aligned}
  \mathcal I_i:~\mathbb{R}^{c_{i-1}\times n_{i-1}^3}
    ~&\longrightarrow~
    \mathbb{R}^{c_i\times n_i^3},\\
  Z^{(i)}&\longmapsto
    \mathrm{ReLU}\bigl(K_i \ast Z^{(i)} + b_i\bigr)
    \triangleq Z^{(i+1)},
\end{aligned}
\] 
where \(K_i\in\mathbb{R}^{c_i\times c_{i-1}\times k_i^3}\) is the convolution kernel, \(b_i\in\mathbb{R}^{c_i}\) is the bias, \(*\) denotes 3D convolution with stride \(s_i\) and padding \(p_i\), and \(\text{ReLU}(\bullet)=\max(0,\bullet)\) is applied element‐wise.  The bottleneck representation after $L=3$ encoding layers is  
\[
Z^{(3)}=(\mathcal I_3\circ\mathcal I_2\circ\mathcal I_1)(X).
\]

\paragraph{Decoder}  
Write \(\hat Z^{(3)}=Z^{(3)}\), each decoder layer \(\mathcal D_i\) perform as:
\[
\begin{aligned}
     \mathcal D_i:~\mathbb{R}^{c_i\times n_i^3}
     ~&\longrightarrow~
     \mathbb{R}^{c_{i-1}\times n_{i-1}^3},\\
    \hat Z^{(i)}&\longmapsto
    \phi_i\bigl(K'_i \star \hat Z^{(i)} + b_i\bigr)
    \triangleq \hat Z^{(i-1)},
\end{aligned}
\] 
where \(K'_i\in\mathbb{R}^{c_{i-1}\times c_i\times k_i^3}\) is the transposed‐convolution kernel; \(b'_i\in\mathbb{R}^{c_{i-1}}\) is the bias;  
\(\star\) denotes 3D transposed convolution with stride \(s_i\), padding \(p_i\), and output‐padding \(s_i-1\); \(\phi_i=\text{ReLU}\) for \(i>1\), and \(\phi_1=\mathrm{Id}\) (no activation) on the final layer.  The reconstruction is  
\[
\hat X
=\hat Z^{(0)}
=(\mathcal D_1\circ\mathcal D_2\circ\mathcal D_3)(\hat Z^{(3)}).
\]

\paragraph{Overall Mapping}  
Compactly, the CAE implements  
\[
\begin{aligned}
\hat X
&=(\mathcal D_1\circ\mathcal D_2\circ\mathcal D_3)
\circ
(\mathcal I_3\circ\mathcal I_2\circ\mathcal I_1)(X),\\
&\mathbb{R}^{3\times48^3}\to\mathbb{R}^{3\times48^3}.
\end{aligned}
\]
 \subsection{Neural Dynamical Transfer Learning framework}\label{sec:ndtl}
\paragraph{Training Data Generation and Augmentation}
As detailed above, each sampled field  
\(\;X\in\mathbb{R}^{3\times48\times48\times48}\)\  
is reshaped to a PyTorch tensor of size \((B,c_0=3,D=48,H=48,W=48)\) before entering the CAE. To mitigate the risk of overfitting due to the limited number of raw ODE trajectories, we augment each sample by  
\begin{enumerate}[(i)]
  \item randomly perturbing the parameters of ordinary differential equation (ODE) by up to \(\pm10\%\), or  
  \item adding zero-mean Gaussian noise of standard deviation \(10^{-2}\).  
\end{enumerate}
These augmentations generate a rich dataset spanning both parameter variations and noise realizations.  
 
We further validate that the CAE model can compress and reconstruct multiple models simultaneously. When pre-training on up to five systems (Lorenz, Lv's, Chen's, the Hastings Powell food chain model, and SEIRS), training remains stable and yields high-fidelity reconstructions in every case.  

\paragraph{CAE Pre-training}
As illustrated above, the composition
\[
\mathcal{D}\circ \mathcal{I}
\;:\;
\mathbb{R}^{3\times48\times48\times48}
\longrightarrow
\mathbb{R}^{3\times48\times48\times48}
\]
reconstructs an input field \(F\) as \(\hat F = \mathcal{D}(\mathcal{I}(F))\). We learn the CAE parameters \(\theta\) by minimizing the mean-squared reconstruction error:
\begin{equation*}
  \mathcal{L}_{\mathrm{MSE}}(\theta)
  = \frac{1}{|\mathcal{B}|}
    \sum_{F\in\mathcal{B}}
    \big\|\,F \;-\;\mathcal{D}\bigl(\mathcal{I}(F)\bigr)\big\|_{2}^{2}
\end{equation*}
over mini-batches \(\mathcal{B}\). Optimization is performed using the Adam algorithm (initial learning rate \(10^{-3}\)), with early stopping as the learning rate falls below \(10^{-6}\) or after a fixed maximum number of epochs.

\paragraph{Transfer Learning Formulation}
After pre-training, the encoder \(\mathcal{I}\) is fixed and used to extract multi-layer latent features.  For any field \(F\), let
\[
F^{\,i} = \mathcal{I}^{i}(F)
\;\in\;
\mathbb{R}^{c_i\times n_i\times n_i\times n_i},
\]
 for $i=1,2,3.$
Denote by $F_i$ the output of the \(i\)-th hidden layer of \(\mathcal{I}\).  We also define the corresponding Gram matrix as:
\begin{equation*}
  G^{\,i}(F)
  = \frac{1}{c_i\,n_i^3}\;
    F^{\,i}\,(F^{\,i})^{T}
  \;\in\;\mathbb{R}^{c_i\times c_i},
\end{equation*}
which encodes style information via channel-wise correlations.

Given two {\it parent} systems \(A\) and \(B\) with the tensor fields \(F_{A}\) and \(F_{B}\), we write $F_{A}^{\,i},\;G_{A}^{\,i}
~\text{and}~
F_{B}^{\,i},\;G_{B}^{\,i}$
for their content and style features at layer \(i\).  We seek a fused field \(\hat F\) whose latent features \(\hat F^{\,i} = \mathcal{I}^{i}(\hat F)\) and whose Gram matrices \(\hat G^{\,i} = G^{\,i}(\hat F)\) satisfy
$
\hat F^{\,i}\approx F_{A}^{\,i},
\
\hat G^{\,i}\approx G_{B}^{\,i}.
$
This is achieved by minimizing the objective function
\begin{equation*}
  \mathcal{L}_{cs}(\hat F)
  = \sum_{i=1}^{N_l}
    \Bigl(
      \alpha_{c}^{i}\,
      \big\|\hat F^{\,i} - F_{A}^{\,i}\big\|_{2}
    + \beta_{s}^{i}\,
      \big\|\hat G^{\,i} - G_{B}^{\,i}\big\|_{2}
    \Bigr)
\end{equation*}
with respect to \(\hat F\).  Here, \(\alpha_{c}^{i},\beta_{s}^{i}\) are nonnegative weights balancing content and style at each layer.  The optimization is carried out by the gradient descent (Adam, initial learning rate \(0.1\)), terminating as \(\mathcal{L}\) falls below a prescribed threshold (e.g.\ \(10^{-6}\)).

For clarity of presentation, in the main text we describe the fusion task in the canonical setting where the \emph{content} of system \(A\) is combined with the \emph{style} of system \(B\). More generally, different fusion schemes can be realized by combining different terms in the optimization objective. For example,
\[
\mathcal{L}_{cc}(\hat F)
  = \sum_{i=1}^{N_l}
    \Bigl(
      \alpha_{c}^{i}\,
      \big\|\hat F^{\,i} - F_{A}^{\,i}\big\|_{2}
    + \beta_{c}^{i}\,
      \big\|\hat F^{\,i} - F_{B}^{\,i}\big\|_{2}
    \Bigr)
\]
corresponds to the setting used in our experiments, where the fused field \(\hat F\) is guided by content features extracted from system \(A\) and system \(B\). This objective is adopted in fusion of the Lorenz system and Lv's system, the Hastings Powell food chain experiment, and the classification induced by fusion using content features. More broadly, the same procedure generalizes naturally to fusing \textit{more than two} parent systems by including additional content and/or style features in \(\mathcal{L}(\hat F)\), with the corresponding weights controlling the relative contributions of each parent.
 \subsection{Coordinate and speed normalization}\label{app:normalization}
\subsubsection{Affine rescaling of phase-space regions}\label{sec:spacenorm}
Different dynamical systems may have attractors supported on distinct subsets of \(\mathbb{R}^n\).  To bring all systems into a common domain and enable their fusion, we apply an axis‐aligned affine map $T$. Under the action of \(T\), any vector field sampled on \(\prod_{i=1}^n[a_i,b_i]\) is rescaled into the unit cube \([0,1]^n\).  

In our experiments where \(n=3\), we compute \(a_i\) and \(b_i\) as the observational minimum and maximum over the training trajectories for each coordinate, and then apply \(T\) to every sample. 

In what follows, we present the general transformation formula and then explicitly illustrate it using the classical Lorenz system.

Consider an $n$ dimensional ODE:
\[
\dot{\mathbf x} = \mathbf F(\mathbf x), 
\qquad
\mathbf x\in\prod_{i=1}^n[a_i,b_i],
\]
and define 
\(\mathbf a=(a_1,\dots,a_n)^\top\) and \(\mathbf b=(b_1,\dots,b_n)^\top\).  
We define the affine map by
\[
T:\mathbb R^n\to\mathbb R^n,\qquad
T(\mathbf x)
= D\,(\mathbf x-\mathbf a)+\mathbf a',
\]
with $D=\mathrm{diag}\Bigl(\tfrac{b'_i - a'_i}{\,b_i - a_i\,}\Bigr)_{i=1}^n$ and $\mathbf a'=(a'_1,\dots,a'_n)^\top,$
which carries \(\prod_{i=1}^n[a_i,b_i]\) onto \(\prod_{i=1}^n[a'_i,b'_i]\).  
Setting \(\mathbf y=T(\mathbf x)\) yields the transformed ODEs:
\[
\dot{\mathbf y}
= D\,\mathbf F\!\bigl(D^{-1}(\mathbf y-\mathbf a')+\mathbf a\bigr).
\]
This transformation ensures the attractor region is accurately mapped into the desired region.

We demonstrate the above method explicitly for the Lorenz system:
\[
\begin{array}{l}
     \dot{x} = \sigma(y - x), \\
     \dot{y} = x(\rho - z) - y, \\
     \dot{z} = xy - \beta z,
\end{array}
\]
with the standard parameters $\sigma=10$, $\rho=28$, and $\beta=\frac{8}{3}$. Its attractor typically resides in the region: $[-25,25]\times[-25,25]\times[0,50].$
We aim to map this attractor into the unit cube $[0,1]^3$. Consider the transformation: $X=\frac{x+25}{50},~Y=\frac{y+25}{50}$, and $Z=\frac{z}{50}.$ The inverse transformation is: $x=50X-25,~y=50Y-25$, and $z=50Z$.
The Lorenz system under the new coordinates $(X,Y,Z)$ becomes:
\[
\begin{array}{l}
     \dot{X} = 10(Y - X), \\
     \dot{Y} = (X - 0.5)(28 - 50Z) - (Y - 0.5),\\
     \dot{Z} = 50(X - 0.5)(Y - 0.5) - \frac{8}{3}Z.
\end{array}
\]
Thus, the attractor of the Lorenz system is now embedded within the unit cube $[0,1]^3$. 
 \subsubsection{System speed normalization}\label{sec:speednorm}
In the phase-space region of interest, different dynamical systems typically evolve with different \emph{average speeds}, which is reflected in the discretized vector fields through the overall magnitude of the sampled vectors. In the fusion procedure, if the two parent systems have substantially different vector field norms, the optimization may be biased toward the faster system, causing the fused dynamics to appear disproportionately similar to the parent with larger speed. When necessary, we mitigate this effect by applying a speed normalization to the discretized vector fields prior to feature extraction and fusion, thereby reducing trivial dominance due to scale differences.

Specifically, given a system \(\dot{\mathbf{x}} = f(\mathbf{x})\) and a prescribed phase-space box \(\mathcal{B}\), we estimate its average speed as the empirical mean of \(\|f(\mathbf{x})\|\) over sampled states \(\mathbf{x}\in\mathcal{B}\). Sampling is performed either on an equispaced Cartesian grid (with \(n_{\text{per-axis}}\) points per dimension) or by uniform random sampling (with \(n_{\text{samples}}\) points), and the speed is computed using either the \(\ell_2\) norm or the \(\ell_1\) norm. Denoting the resulting estimate by \(v_{\mathrm{avg}} \approx \mathbb{E}_{\mathbf{x}\in \mathcal{B}}\big[\|f(\mathbf{x})\|\big]\), we rescale the system by multiplying all coefficients in the governing equations by a single factor \(s = v_{\mathrm{target}}/v_{\mathrm{avg}}\), where \(v_{\mathrm{target}}>0\) is a prescribed reference speed. This operation produces a new vector field \(f_{\mathrm{scaled}}(\mathbf{x}) = s\,f(\mathbf{x})\), equivalently a uniform time-rescaling of the dynamics, and ensures that the average speed within the region of interest matches \(v_{\mathrm{target}}\). 

In our implementation, we first apply the phase-space normalization described in Sec.~\ref{app:normalization} to map the parent attractors into \(\mathcal{B}=[0,1]^n\), and then perform the above speed normalization on this common domain.
 \FloatBarrier
\section{\codexnew{Transformation Properties of Content and Style Features}}
\label{app:style_robustness}
\codexnew{We evaluate how the input-level content tensor and Gram representation respond to physically meaningful transformations applied at the ODE level. For each rescaled baseline system \((\dot{\mathbf{x}}=f(\mathbf{x}),\ \mathbf{x}=(x,y,z))\), the vector field is sampled on a fixed grid to obtain \(V\in\mathbb{R}^{D\times H\times W\times 3}\). We then apply time reversal \((f\mapsto-f)\), coordinate exchange \((x\leftrightarrow y)\), reflection \((x\mapsto-x)\), and rotational conjugacy about the \(z\)-axis.}

\codexnew{The normalized pointwise discrepancy, denoted by \(\delta_{\mathrm{point}}\), is the mean-squared difference between \(V\) and the transformed tensor \(V_T\), normalized by \(\mathbb{E}\|V\|^2\). The Gram discrepancy, \(\delta_{\mathrm{Gram}}\), is the mean-squared difference between their Gram matrices after channel-wise standardization removes pure scale effects. For coordinate exchange, the corresponding channel permutation is aligned before comparison so that the reported discrepancy does not arise solely from renaming the vector field components.}

\codexnew{Table~\ref{table_style_exp} shows that time reversal produces a large pointwise change while leaving the Gram representation exactly unchanged. Coordinate exchange likewise yields a vanishing Gram discrepancy after channel alignment. Reflection and rotation do alter the Gram representation, but less than they alter the pointwise tensor for all four systems considered. Overall, the Gram representation is invariant under time reversal and coordinate exchange after channel alignment, and remains comparatively stable under reflection and rotation for the four systems considered.}

\begin{table}[htb]
\codexnewcolor
\centering
\caption{Transformation-induced discrepancies of the input-level content tensor and Gram representation. Both measures are dimensionless; smaller values indicate greater stability under the specified transformation.}
\label{table_style_exp}
\footnotesize
\renewcommand{\arraystretch}{1.08}
\setlength{\tabcolsep}{2.5pt}
\begin{tabular}{llcc}
\toprule
System & Transformation & $\delta_{\mathrm{point}}$ & $\delta_{\mathrm{Gram}}$ \\
\midrule
Lorenz & Time reversal & 4.000 & 0.000 \\
Lorenz & Coordinate exchange $x\leftrightarrow y$ & 1.033 & $2.76\times10^{-31}$ \\
Lorenz & Reflection $x\mapsto -x$ & 12.543 & 0.115 \\
Lorenz & Rotation $R_z(30^\circ)$ & 1.734 & 0.119 \\
\midrule
Chen & Time reversal & 4.000 & 0.000 \\
Chen & Coordinate exchange $x\leftrightarrow y$ & 3.670 & $4.61\times10^{-31}$ \\
Chen & Reflection $x\mapsto -x$ & 8.804 & 0.695 \\
Chen & Rotation $R_z(30^\circ)$ & 1.852 & 0.270 \\
\midrule
Lv & Time reversal & 4.000 & 0.000 \\
Lv & Coordinate exchange $x\leftrightarrow y$ & 3.588 & $4.16\times10^{-30}$ \\
Lv & Reflection $x\mapsto -x$ & 8.223 & 0.654 \\
Lv & Rotation $R_z(30^\circ)$ & 1.932 & 0.275 \\
\midrule
R\"ossler & Time reversal & 4.000 & 0.000 \\
R\"ossler & Coordinate exchange $x\leftrightarrow y$ & 1.073 & $7.42\times10^{-26}$ \\
R\"ossler & Reflection $x\mapsto -x$ & 7.989 & 0.258 \\
R\"ossler & Rotation $R_z(30^\circ)$ & 0.336 & 0.0856 \\
\bottomrule
\end{tabular}
\end{table}
 \FloatBarrier
\section{\codexnew{Closed-Form Recovery and Dynamical Analysis}}\label{app:numerics}
\subsection{Regression techniques for closed-form vector field recovery}\label{sec:sindy}
To recover a closed‐form representation of the underlying vector field from discrete vector field, we employ the regression technique inspired by~\cite{celso_sindy2011PRL, celso_sindy2011PRX}. Denote by
\[
\mathbf{f}
\;=\;
\begin{pmatrix}
f(x_1)^\top\\
\vdots\\
f(x_N)^\top
\end{pmatrix}
\;\in\;\mathbb{R}^{N\times d}
\]
the collection of observational velocity vectors at \(N\) grid points \(x_j\in\mathbb{R}^d\).  We build a dictionary of \(p\) candidate basis functions
\(\{\varphi_k(x)\}_{k=1}^p\) (e.g., monomials up to second-order, plus problem‐specific nonlinear terms), and assemble the design matrix as:
\[
\mathbf{X}
=
\begin{pmatrix}
\varphi_1(x_1) & \varphi_2(x_1) & \cdots & \varphi_p(x_1)\\
\varphi_1(x_2) & \varphi_2(x_2) & \cdots & \varphi_p(x_2)\\
\vdots         & \vdots         &        & \vdots         \\
\varphi_1(x_N) & \varphi_2(x_N) & \cdots & \varphi_p(x_N)
\end{pmatrix}
\;\in\;\mathbb{R}^{N\times p}.
\]
We seek the coefficient matrix \(\Theta\in\mathbb{R}^{p\times d}\) satisfying
\[
\mathbf{f}
\;=\;
\mathbf{X}\,\Theta
\;+\;
\boldsymbol{\epsilon},
\]
where \(\boldsymbol{\epsilon}\) denotes the residual.

\subsubsection{General Regularized Regression}
We fit \(\Theta\) by solving the convex program
\[
\hat\Theta
\;=\;
\arg\min_{\Theta}
\;
\underbrace{\bigl\|\mathbf{f}-\mathbf{X}\Theta\bigr\|_{F}^{2}}_{\text{data fidelity}}
\;+\;
\underbrace{\lambda\,\mathcal{R}(\Theta)}_{\substack{\text{regularization}}}\,,
\]
where \(\|\cdot\|_F\) is the Frobenius norm, \(\lambda\ge0\) is a tuning parameter, and
\[
\mathcal{R}(\Theta)
=
\begin{cases}
0, & \text{(Linear)},\\[6pt]
\|\Theta\|_{1,1} = \sum\limits_{k,\ell}|\Theta_{k\ell}|, & \text{(Lasso)},\\[6pt]
\|\Theta\|_{F}^{2}, & \text{(Ridge)}.
\end{cases}
\]
In practice, we also allow an optional binary mask \(M\in\{0,1\}^{p\times d}\) to enforce known sparsity patterns by replacing \(\mathbf{X}\) with \(\mathbf{X}\odot M\) or zeroing out disallowed entries of \(\Theta\) after each iteration.

The optimization result \(\hat\Theta\) defines the closed-form vector field approximation
$
\hat f(x)
=\sum_{k=1}^{p}\hat\Theta_{k,:}\,\varphi_k(x).
$

\subsubsection{System‐specific Basis and Regression Method}
Experiments indicate that the system reconstructed via Linear regression more faithfully preserves the geometric structure of the parent system—its attractor is retained with high fidelity. However, a large number of nonlinear terms undermines the interpretability of the resulting vector field. In contrast, the system obtained by the Lasso regression captures only the most critical nonlinear terms, yielding a vector field with greater physical significance. Specifically, the dictionary of the basis functions and the regression methods used in the experiments reported in the main text are as follows.
\begin{itemize}
  \item Fusion among Lorenz‐like system and classification induced by fusion using content features:
  \begin{itemize}
      \item Basis = all monomials of total degree \(\le2\),
      \item Regression mode: Linear (\(\lambda=0\));
  \end{itemize}

  \item Ecological creation in the Hastings Powell food chain model:
  \begin{itemize}
      \item Model family = the Hastings-Powell food chain equations in Appendix~\ref{app:ecology},
      \item Regression mode: constrained least squares after an outer search over the two saturation parameters;
  \end{itemize}
  
  \item Fusion of the SEIRS system and the Lorenz system:
  \begin{itemize}
      \item Basis = all monomials of total degree \(\le2\),
      \item Regression mode: Lasso (\(\mathcal{R}(\Theta)=\|\Theta\|_{1,1}\)), with \(\lambda\) chosen by cross‐validation.
  \end{itemize}
\end{itemize}
 \subsection{Numerical evaluation of dynamical signatures}\label{sec:num}
Below we describe the numerical procedures used to compute the dynamical signatures reported in the main text. 
\subsubsection{The Lyapunov spectrum}
To compute the Lyapunov spectrum for an ODE \(\dot{\mathbf{x}}=f(\mathbf{x})\), we generate a trajectory \(\{\mathbf{x}_k\}_{k=0}^{T}\) from a prescribed initial condition using a fixed time step \(\Delta t=h\) (after optionally omitting an initial transient). Along this trajectory, we estimate the Jacobian \(J_k=\nabla f(\mathbf{x}_k)\) by finite differences applied to the right-hand side \(f\). We then evolve an orthonormal basis \(U_k\in\mathbb{R}^{d\times d}\) of the tangent space using a backward-Euler discretization of the variational equation,
\[
U_{k+1} = \bigl(I - J_k \Delta t\bigr)^{-1} U_k,
\]
followed by a QR decomposition \(U_{k+1}=Q_k R_k\). The logarithms of the absolute diagonal entries of \(R_k\) are accumulated over time, and the Lyapunov exponents are obtained as time-averaged growth rates,
\[
\lambda_j \approx \frac{1}{T\,\Delta t}\sum_{k=1}^{T} \log\bigl| (R_k)_{jj} \bigr|,\qquad j=1,\dots,d,
\]
with the resulting spectrum reported in decreasing order. In practice, the iteration is terminated either after a prescribed trajectory length or when the QR updates become numerically uninformative according to a stopping heuristic based on tolerance.

\subsubsection{The Kaplan-Yorke dimension}
The Kaplan-Yorke dimension is computed directly from the Lyapunov spectrum \(\{\lambda_i\}_{i=1}^{d}\), sorted in decreasing order \(\lambda_1\ge \lambda_2\ge \cdots \ge \lambda_d\). Let \(S_k=\sum_{i=1}^{k}\lambda_i\) be the cumulative sum. We identify the largest index \(j\) such that \(S_j\ge 0\) (if no such index exists, we set \(D_{\mathrm{KY}}=0\)). The Kaplan-Yorke dimension is then given by
\[
D_{\mathrm{KY}} \;=\; j \;+\; \frac{S_j}{|\lambda_{j+1}|},
\]
with a small numerical constant added in the denominator to avoid division by zero. In implementation, when the cumulative sum does not cross zero within the available spectrum (which can indicate ill-posedness or undersampling), we cap \(j\le d-2\) to ensure \(\lambda_{j+1}\) is defined.

\subsubsection{The intrinsic dimension}
We estimate the intrinsic dimension of the attractor from the trajectory point cloud using the TwoNN estimator, with an implementation adapted from the opensource codebase associated with ~\cite{gilpin_benchmark_nips2021}. Given a simulated trajectory \(\{\mathbf{x}_t\}\), we first discard an initial transient and optionally downsample by taking \(\mathbf{x}_{t}\) with a fixed stride, yielding a point set \(X=\{\mathbf{x}_n\}_{n=1}^{N}\subset\mathbb{R}^{d}\) (with \(N\gtrsim 20\) for numerical stability). For each point \(\mathbf{x}_n\), we compute the Euclidean distances \(r_1(\mathbf{x}_n)\) and \(r_2(\mathbf{x}_n)\) to its first and second nearest neighbors, respectively, and form the ratio \(\mu_n = r_2(\mathbf{x}_n)/r_1(\mathbf{x}_n)\). For time series we employ a Theiler window \(w\) to mitigate temporal correlations: when selecting nearest neighbors for \(\mathbf{x}_n\), we exclude candidate points \(\mathbf{x}_m\) with \(|m-n|\le w\), so that \(r_1(\mathbf{x}_n)\) and \(r_2(\mathbf{x}_n)\) are computed from spatial neighbors rather than trivially adjacent samples along the trajectory. The intrinsic dimension is then estimated by
\[
D_{\mathrm{ID}}
\;=\;
\left(\frac{1}{M}\sum_{n=1}^{M}\log \mu_n\right)^{-1}
\;=\;
\frac{M}{\sum_{n=1}^{M}\log \mu_n},
\]
where the sum is taken over the \(M\) valid samples with finite \(\mu_n>1\). Nearest-neighbor queries are performed via exact \(k\)-NN search (and optionally via approximate search for acceleration), and we record failures when no valid neighbor pairs remain after Theiler exclusion.


\subsubsection{The invariant measure statistics}
To quantify how closely a created system matches a reference system in terms of invariant measure statistics, we approximate the invariant measure by the empirical distribution of long-time trajectory samples and compute a divergence between the resulting distributions. For each system and each prescribed initial condition, we simulate trajectories over a fixed time window with step size \(\Delta t\), discard an initial transient by removing samples with \(t<t_{\mathrm{tr}}\), and retain the remaining point cloud in phase space. For an \(n\) dimensional system, we construct a common \(n\) dimensional histogram on a shared bounding box \(\mathcal{B}\subset\mathbb{R}^n\). In practice, since all systems are first normalized as described in Sec.~\ref{app:normalization}, we take \(\mathcal{B}=[0,1]^n\) for divergence computations.

We discretize each coordinate axis of \(\mathcal{B}\) into \(B\) equal-width bins, yielding an \(n\) dimensional histogram with \(B^n\) cells. This produces discrete probability masses \(P\) and \(Q\) after adding a small \(\varepsilon\) to each cell for numerical stability and normalizing to unit sum. We report either the Kullback-Leibler divergence \(\mathrm{KL}(P\|Q)=\sum P\log(P/Q)\) or the symmetric Jensen-Shannon divergence
\[
\mathrm{JS}(P,Q)=\tfrac12\,\mathrm{KL}(P\|M)+\tfrac12\,\mathrm{KL}(Q\|M),\quad
M=\tfrac12(P+Q),
\]
which is numerically more stable. Divergences are computed independently for multiple initial conditions and summarized by their mean (and standard deviation) across initializations.

\subsubsection{Power Spectral Density}
The power spectral density (PSD) is computed to analyze the frequency domain properties of dynamical systems~\cite{fft_psd_book2007}. Given a numerical solution of the ODEs, we first generate a time series $x(t_i)$ sampled uniformly with time step $\Delta t$. The discrete Fourier transform of the series is then computed as:
\[
    X(f)=\sum_{k=0}^{N-1}x(t_k)\mathrm{e}^{-2\pi \mathrm{i}ft_k},\quad f\in[0,f_{\mathrm{Nyquist}}],
\]
where $f_{\mathrm{Nyquist}}=\frac{1}{2\Delta t}$ is the Nyquist frequency. The PSD is obtained as the magnitude squared of the DFT: $P(f)=|X(f)|^2.$
We retain only positive frequencies to interpret physical phenomena clearly. Such spectral analysis enables identification of characteristic frequencies and detection of dynamical regimes within the system.
 \subsection{Dynamical analysis of generated closed-form systems}
\label{app:closed_form}
\label{sec:analysis}
Our pipeline performs symbolic regression or model family projection after optimizing the discretized vector field. This step yields a closed form vector field, which allows standard dynamical analysis to be carried out on the created systems. We first compare bifurcation behavior for a generated Lorenz like system. We then analyze equilibria and stability for one representative member of the SEIRS and Lorenz fusion family described in Sec.~\ref{sec:extra_domain_fusion} and for a fused system created from the Nos\'e Hoover and Rucklidge systems.

\subsubsection{Bifurcation diagram analysis}
\begin{figure}[h!]
\centering
\includegraphics[width=0.48\textwidth]{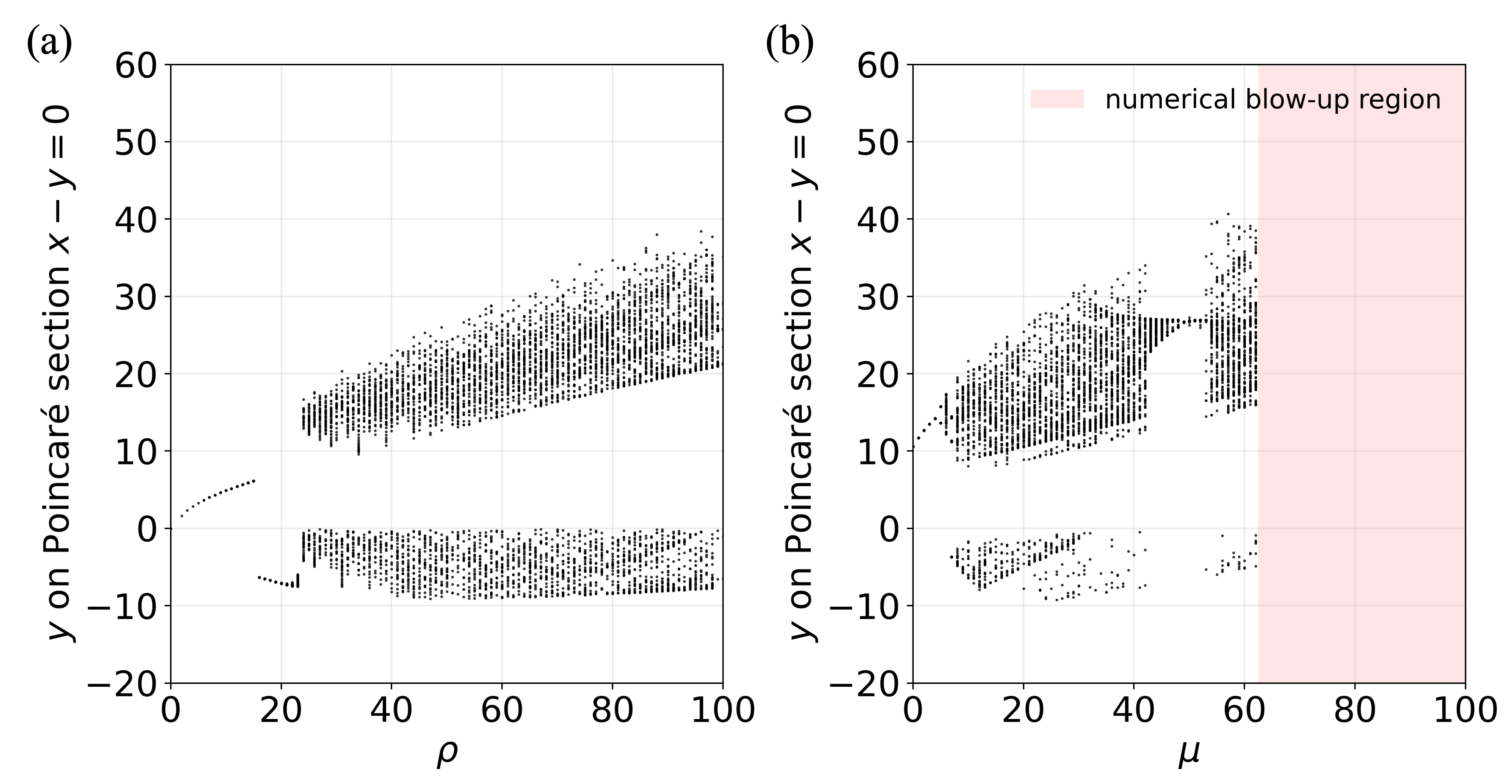}
\caption{Bifurcation diagrams of (a) the classical Lorenz system and (b) the fused system.}
\label{bifurcation}
\end{figure}

To compare bifurcation behavior, we consider the classical Lorenz system and the system obtained by fusing the content feature of the Lorenz and Lv's system at fusion parameter $\frac{\alpha_c}{\alpha_c+\beta_c}=0.5$. For a consistent comparison with the classical Lorenz system, we choose as the bifurcation parameter of the fused system the coefficient $\mu$ of the linear $x$ term in its second equation, i.e., the term playing the role analogous to $\rho x$ in the Lorenz system. For both systems, the bifurcation diagrams are constructed from the same Poincaré section \(x-y=0\), recording only crossings from \(x-y<0\) to \(x-y>0\) and plotting the corresponding \(y\) coordinates. As shown in Fig.~\ref{bifurcation}, the two diagrams differ clearly over most of the parameter range, and are only similar in a limited regime.

\FloatBarrier

\subsubsection{A representative SEIRS and Lorenz fusion}
A fused epidemiological model obtained by fusing the content of an SEIRS system with the style of the Lorenz system and then applying LASSO regression takes the form
\begin{align*}
    \dot{S} &= 0.15R - 0.9SI,\\
    \dot{E} &= -0.1E + 0.9SI,\\
    \dot{I} &= 0.1E - 0.2I,\\
    \dot{R} &= 0.2I - 0.15R.
\end{align*}

together with the conserved total population \(S+E+I+R=1\). Since the right hand sides are polynomial, the vector field is locally Lipschitz continuous. The system has two equilibria: the disease free equilibrium (DFE) \((S,E,I,R)=(1,0,0,0)\) and an endemic equilibrium (EE)
\[
(S^*,E^*,I^*,R^*) = \left( \frac{2}{9},\; \frac{14}{39},\; \frac{7}{39},\; \frac{28}{117} \right).
\]
Linearization at the DFE together with the next generation matrix method yields \(\mathcal{R}_0 = 4.5>1\), indicating instability of the DFE. At the EE, the nonzero Jacobian eigenvalues satisfy
\[
\lambda^3 + 0.61154\lambda^2 + 0.117693\lambda + 0.0105001 = 0,
\]
which, by the Routh Hurwitz criterion, has all roots with negative real parts; hence the EE is locally asymptotically stable.

\subsubsection{A representative Nos\'e Hoover and Rucklidge fusion}
The system obtained by fusing the content of the Nos\'e Hoover and Rucklidge systems and then applying linear regression is
\begin{center}
\resizebox{0.98\linewidth}{!}{$
\begin{aligned}
    \dot{x} &= -4.389 - 0.77x + 10.221y + 5.936z + 0.355x^{2} - 0.185xy - 0.256xz - 1.664y^{2} - 12.071yz + 0.103z^{2}, \nonumber \\
    \dot{y} &= 9.655 - 3.441x - 15.424y - 15.762z - 0.104x^{2} + 0.178xy + 0.136xz - 0.602y^{2} + 31.415yz - 0.149z^{2}, \nonumber \\
    \dot{z} &= -4.566 + 0.218x + 20.505y - 1.011z + 0.364x^{2} + 0.125xy - 0.418xz - 20.319y^{2} - 0.146yz + 0.984z^{2}, \nonumber
\end{aligned}
$}
\end{center}
where \((x,y,z)\in\mathbb{R}^{3}\). The right hand sides are quadratic polynomials and hence continuously differentiable on \(\mathbb{R}^3\); consequently, the vector field is locally Lipschitz on any compact domain. A direct computation of the divergence of \(\mathbf{F}=(\dot{x},\dot{y},\dot{z})\) yields
\[
\begin{aligned}
\nabla \cdot \mathbf{F}
&= \frac{\partial \dot{x}}{\partial x}
 + \frac{\partial \dot{y}}{\partial y}
 + \frac{\partial \dot{z}}{\partial z}\\
&= -17.205 + 0.47x - 1.535y + 33.127z,
\end{aligned}
\]

which is not identically zero and depends on the state variables, indicating that the system is not conservative in general. We compute the Lyapunov spectrum using the QR based algorithm described in Appendix~\ref{app:numerics} along a numerically integrated trajectory. For the initial condition \((x_0,y_0,z_0)=(0.3,0.3,0.5)\), we obtain
\[
(\lambda_1,\lambda_2,\lambda_3) \approx (0.2360,\,-0.2960,\,-0.7950),
\]
where \(\lambda_1>0\) provides numerical evidence of chaotic dynamics. Moreover,
\[
\lambda_1+\lambda_2+\lambda_3 \approx -0.8549 < 0,
\]
which indicates average phase space volume contraction along this trajectory and is consistent with dissipative behavior.

We further compute equilibrium points by solving \(\mathbf{F}(x,y,z)=\mathbf{0}\) and obtain four real equilibria (Newton refined; \(\|\mathbf F(x^*,y^*,z^*)\|\lesssim 10^{-14}\) in double precision):
\begin{center}
\begin{tabular}{c | c c c}
Equilibrium & $x^{*}$ & $y^{*}$ & $z^{*}$ \\ \hline
$E_{1}$ & $7.765915684$ & $-0.619136530$ & $-0.436383769$ \\
$E_{2}$ & $0.605638724$ & $0.656065581$ & $0.572651667$ \\
$E_{3}$ & $0.269388119$ & $0.370912522$ & $0.702402061$ \\
$E_{4}$ & $7.484826042$ & $1.493687675$ & $1.386092068$
\end{tabular}
\end{center}
Let \(\mathbf F(x,y,z)=(\dot x,\dot y,\dot z)^\top\) and \(J=\nabla \mathbf F\). Then
\[
\resizebox{0.98\linewidth}{!}{$
J(x,y,z)=
\begin{pmatrix}
-0.77 + 0.71x - 0.185y - 0.256z & 10.221 - 0.185x - 3.328y - 12.071z & 5.936 - 0.256x - 12.071y + 0.206z\\
-3.441 - 0.208x + 0.178y + 0.136z & -15.424 + 0.178x - 1.204y + 31.415z & -15.762 + 0.136x + 31.415y - 0.298z\\
0.218 + 0.728x + 0.125y - 0.418z & 20.505 + 0.125x - 40.638y - 0.146z & -1.011 - 0.418x - 0.146y + 1.968z
\end{pmatrix}.
$}
\]
Evaluating \(J\) at each real equilibrium yields the following eigenvalues (rounded to \(4\) decimals):
\begin{align*}
\sigma(J(E_1)) &\approx \{\, 2.3822,\; -14.7214 \pm 37.8870\, i \,\},\\
\sigma(J(E_2)) &\approx \{\, -1.5430,\; 1.2929 \pm 5.8862\, i \,\},\\
\sigma(J(E_3)) &\approx \{\, -1.5635,\; 3.5921 \pm 4.3025\, i \,\},\\
\sigma(J(E_4)) &\approx \{\, 1.7325,\; 14.1023 \pm 32.8109\, i \,\}.
\end{align*}
Hence each equilibrium has at least one eigenvalue with \(\Re(\lambda)>0\), and therefore all real equilibria are unstable. Overall, this fused system is dissipative, has unstable equilibria, and exhibits chaotic behavior, consistent with the numerical results.
 \FloatBarrier
\section{\codexnew{Fusion Results across Canonical Chaotic Systems}}\label{app:canonical}
We perform fusions using content features over a collection of canonical chaotic systems, including the Lorenz, Chen's, Lv's, R\"ossler, Hadley, Halvorsen, Nos\'e-Hoover, and Rucklidge systems. The system definitions and their standard parameter settings are taken from the compilation in~\cite{gilpin_benchmark_nips2021}. The experimental results are summarized below.

\begin{figure}[h!]
\centering
\includegraphics[width=0.48\textwidth]{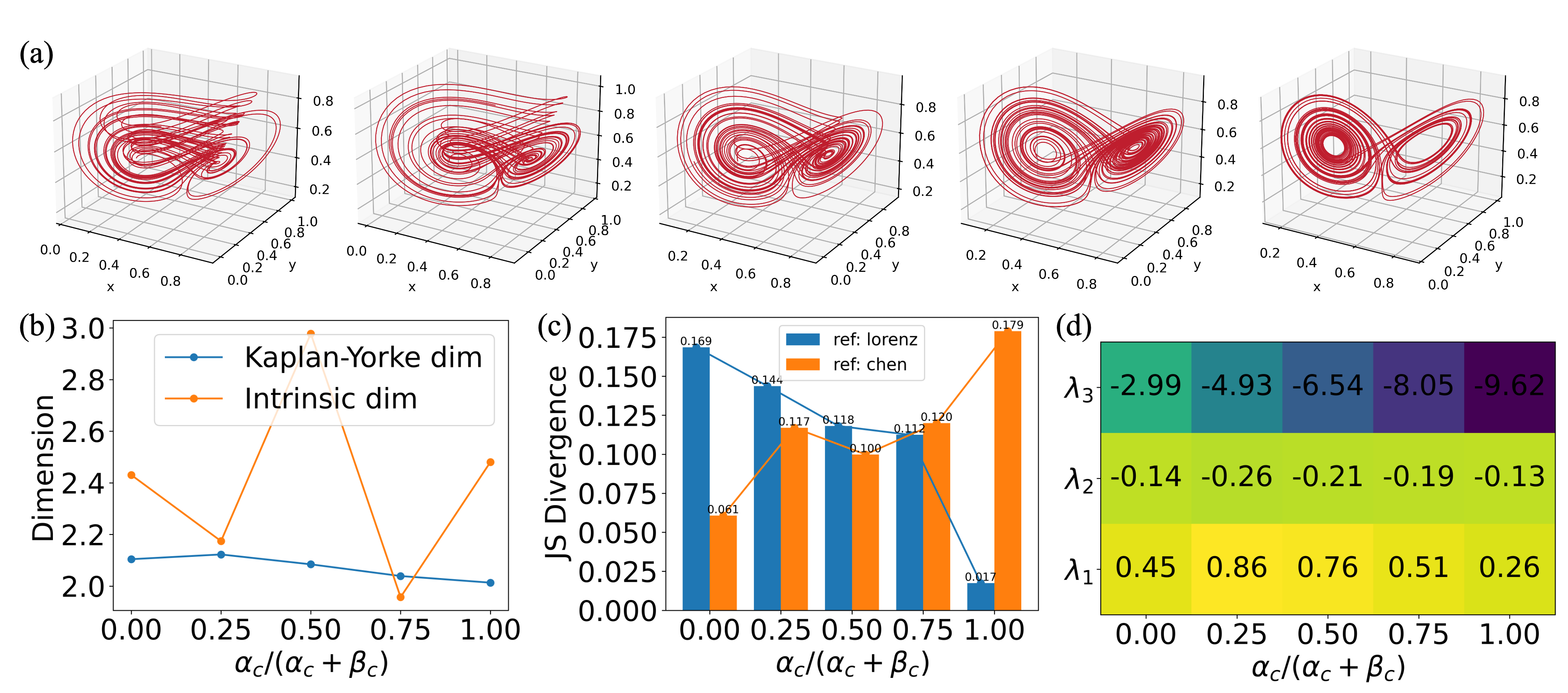}
\caption{Fusion results for the Lorenz and Chen's systems as the content weights change. Initial point used in numerical evaluation is $(0.3,0.3,0.5)$. (a) Phase-space trajectories of the fused systems. From left to right, the fusion parameter $\alpha_c/(\alpha_c+\beta_c)$ is 0.00, 0.25, 0.50, 0.75, and 1.00. (b) Intrinsic and Kaplan-Yorke dimensions of the fused systems. (c) Jensen-Shannon divergence between the invariant measure distribution of the fused system and that of the parent system. (d) Lyapunov spectra of the fused systems.}
\label{lorenz_chen_exp}
\end{figure}

\begin{figure}[h!]
\centering
\includegraphics[width=0.48\textwidth]{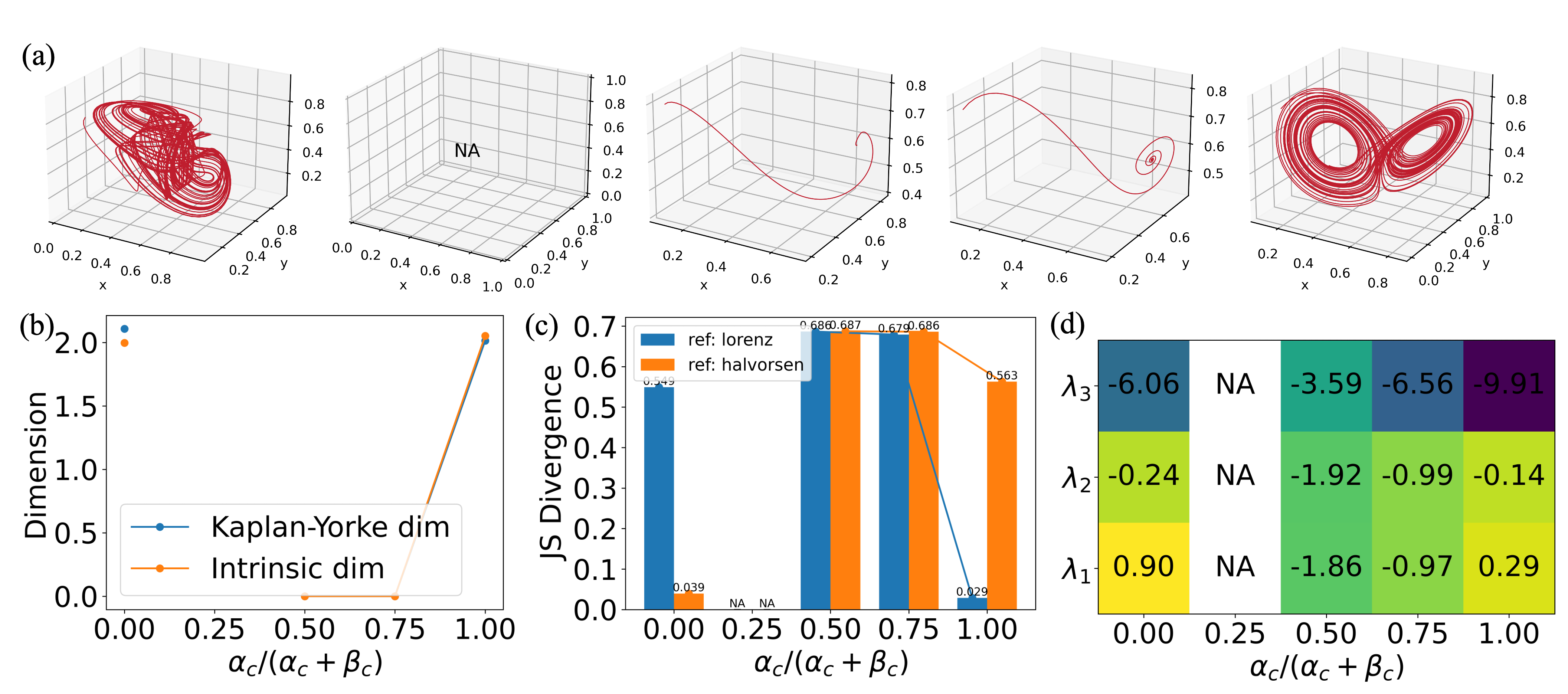}
\caption{Fusion results for the Lorenz and Halvorsen systems as the content weights change. Initial point used in numerical evaluation is $(0.1,0.2,0.8)$. Panels (a)-(d) have the same meanings as in Fig.~\ref{lorenz_chen_exp}.}
\label{lorenz_halvorsen_exp}
\end{figure}

\begin{figure}[h!]
\centering
\includegraphics[width=0.48\textwidth]{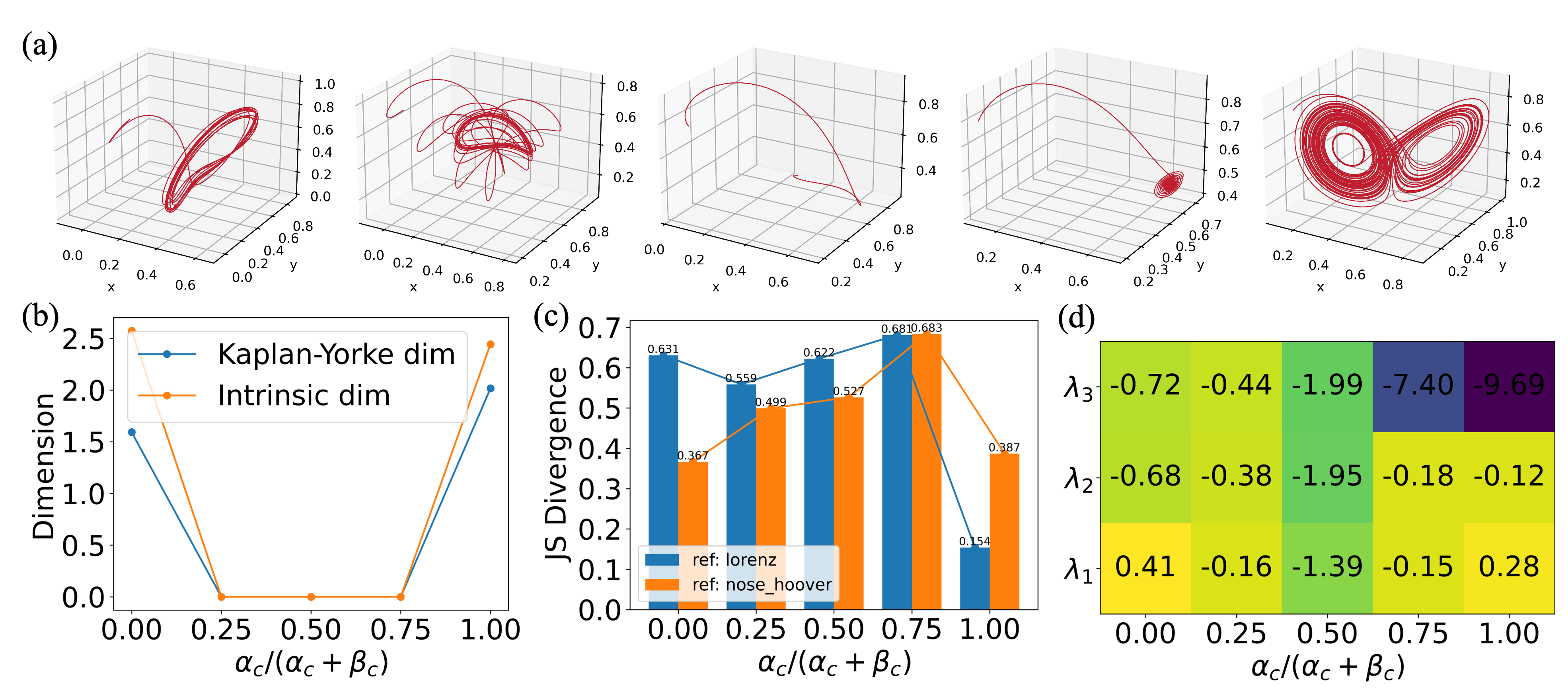}
\caption{Fusion results for the Lorenz and Nos\'e-Hoover systems as the content weights change. Initial point used in numerical evaluation is $(0.1,0.2,0.8)$. Panels (a)-(d) have the same meanings as in Fig.~\ref{lorenz_chen_exp}.}
\label{lorenz_nose_hoover_exp}
\end{figure}

\begin{figure}[h!]
\centering
\includegraphics[width=0.48\textwidth]{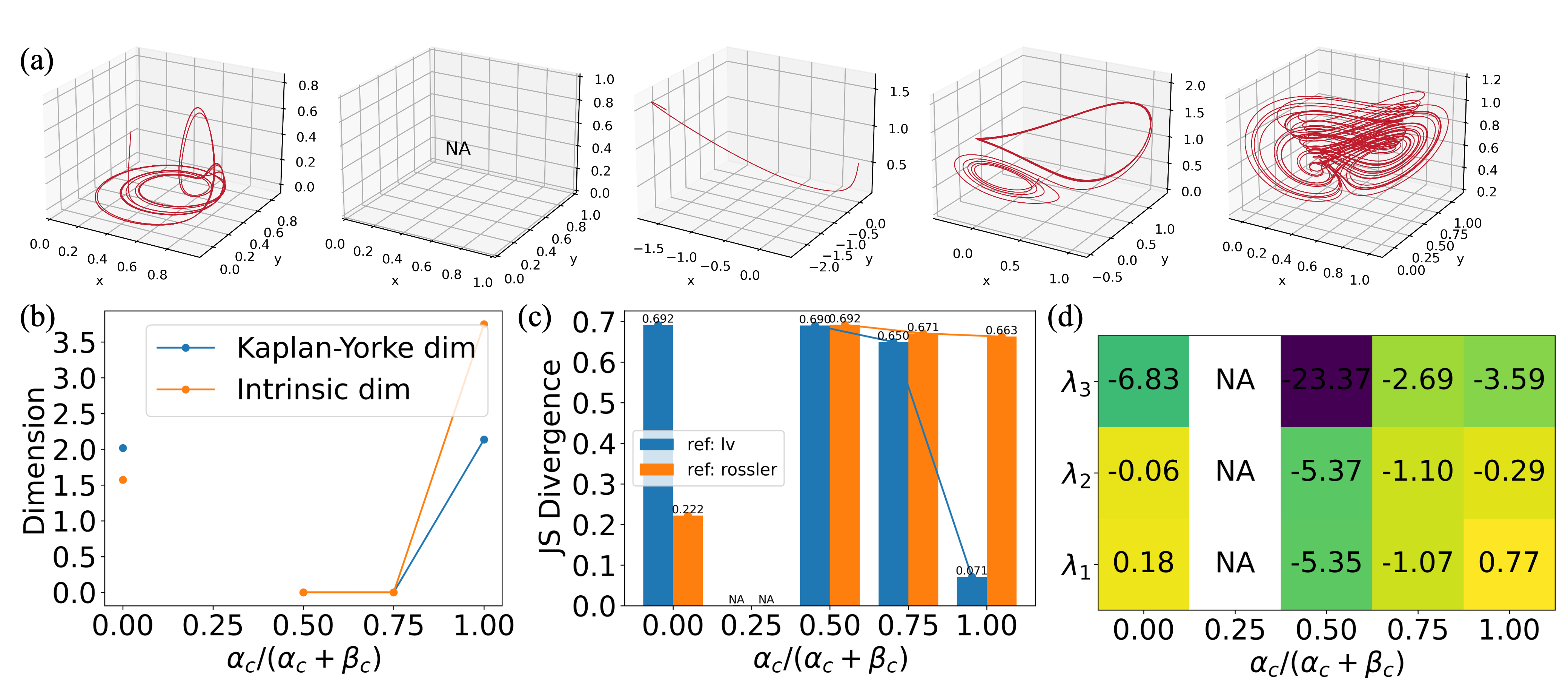}
\caption{Fusion results for Lv's and the R\"ossler systems as the content weights change. Initial point used in numerical evaluation is $(0.3,0.3,0.5)$. Panels (a)-(d) have the same meanings as in Fig.~\ref{lorenz_chen_exp}.}
\label{lv_rossler_exp}
\end{figure}

\begin{figure}[h!]
\centering
\includegraphics[width=0.48\textwidth]{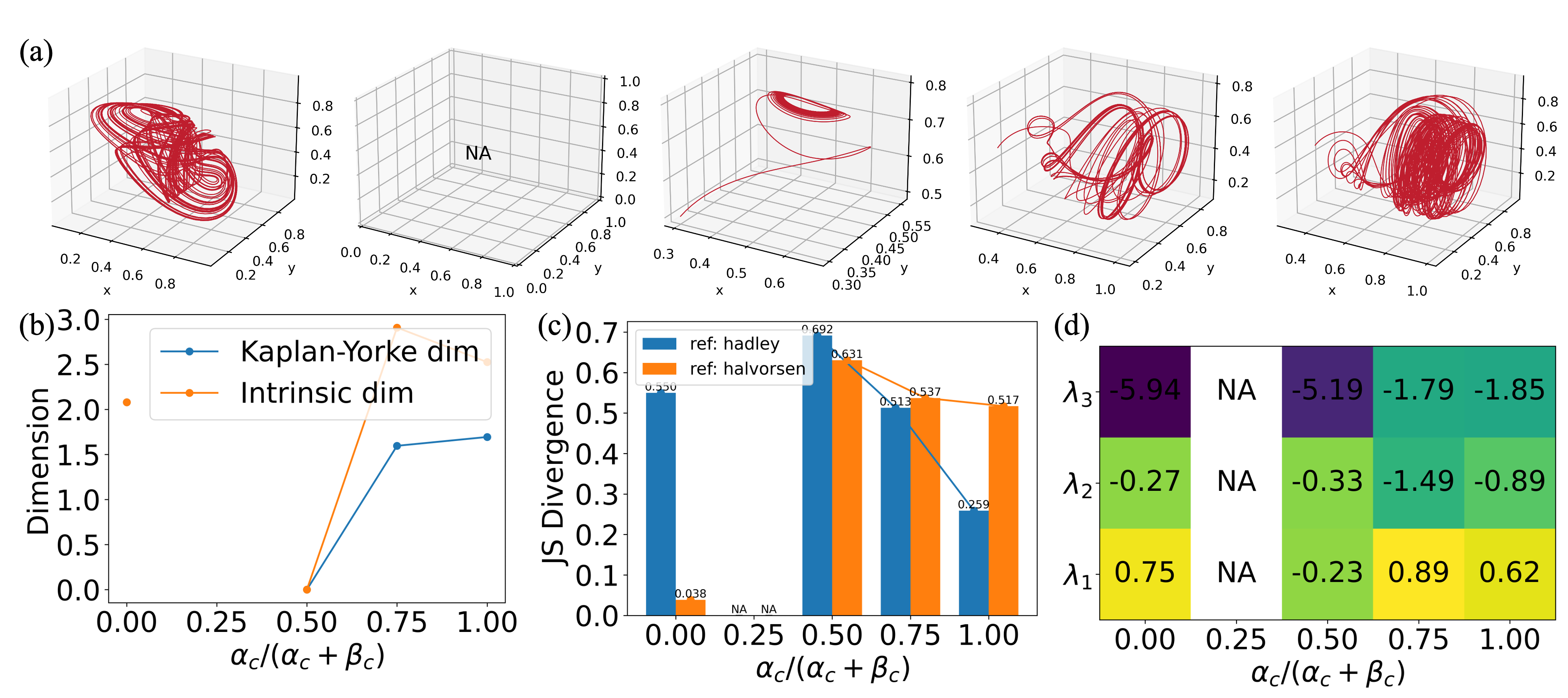}
\caption{Fusion results for the Hadley and Halvorsen systems as the content weights change. Initial point used in numerical evaluation is $(0.1,0.2,0.8)$. Panels (a)-(d) have the same meanings as in Fig.~\ref{lorenz_chen_exp}.}
\label{hadley_halvorsen_exp}
\end{figure}

\begin{figure}[h!]
\centering
\includegraphics[width=0.48\textwidth]{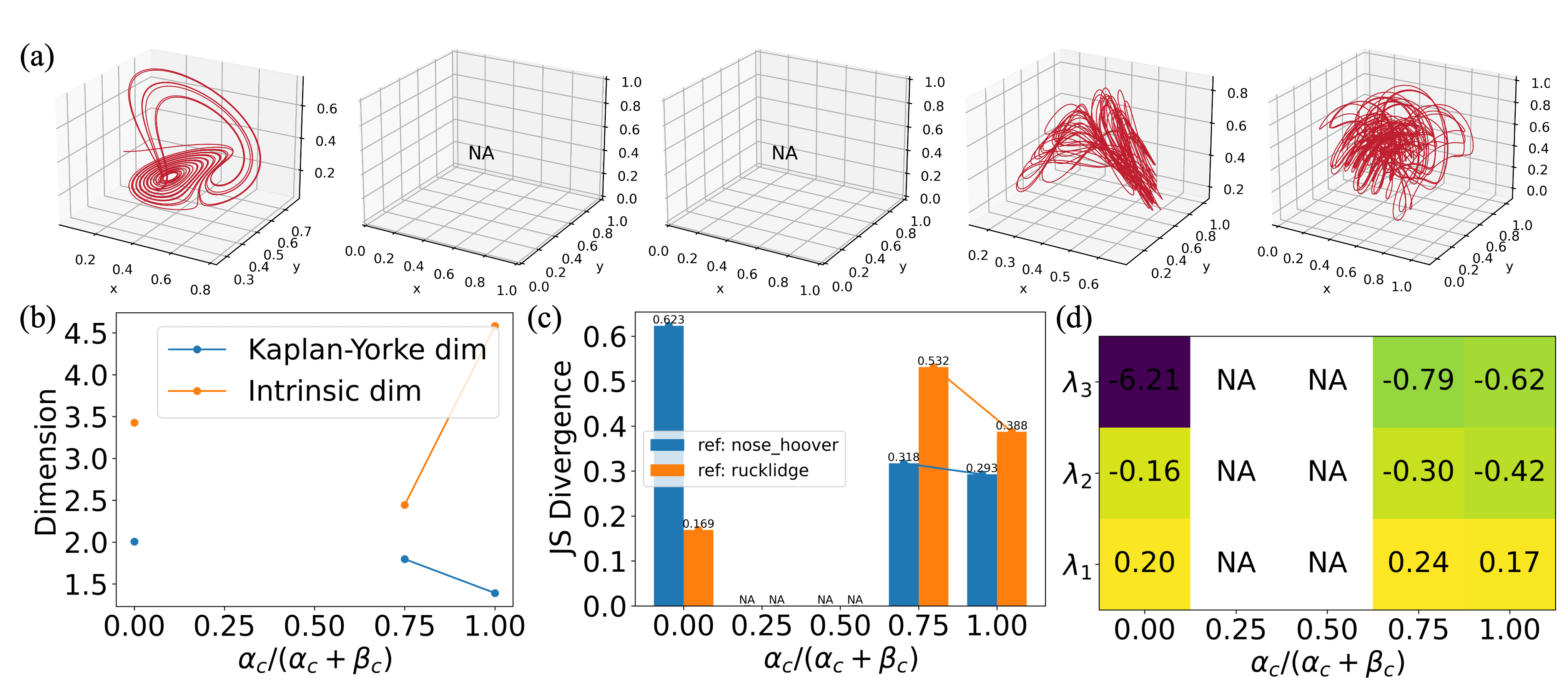}
\caption{Fusion results for the Nos\'e-Hoover and Rucklidge systems as the content weights change. Initial point used in numerical evaluation is $(0.3,0.3,0.5)$. Panels (a)-(d) have the same meanings as in Fig.~\ref{lorenz_chen_exp}.}
\label{nose_hoover_rucklidge_exp}
\end{figure}

Here, entries marked as \texttt{NA} indicate cases where the corresponding trajectory exhibits finite time blowup. By inspecting the Lyapunov spectrum together with the divergence between the invariant measure distributions of the fused and reference systems, we find that certain fusion settings yield nontrivial dynamics and, in particular, create new chaotic systems. Additional experiments can be reproduced using the code repository described in the Data and Code Availability statement.
 \par
\codexnew{Since our fusion procedure operates in the low-dimensional feature space learned by the CAE, it is useful to view feature extraction and fusion from a manifold perspective. The discretized vector field tensors are structured high-dimensional objects, and this structure motivates the use of a CAE to obtain compact latent representations. This interpretation is intended only as a conceptual picture: we do not explicitly identify a latent manifold or assume that every fused field remains on one.}

\begin{figure}[h!]
\codexnewcolor
\centering
\includegraphics[width=0.48\textwidth]{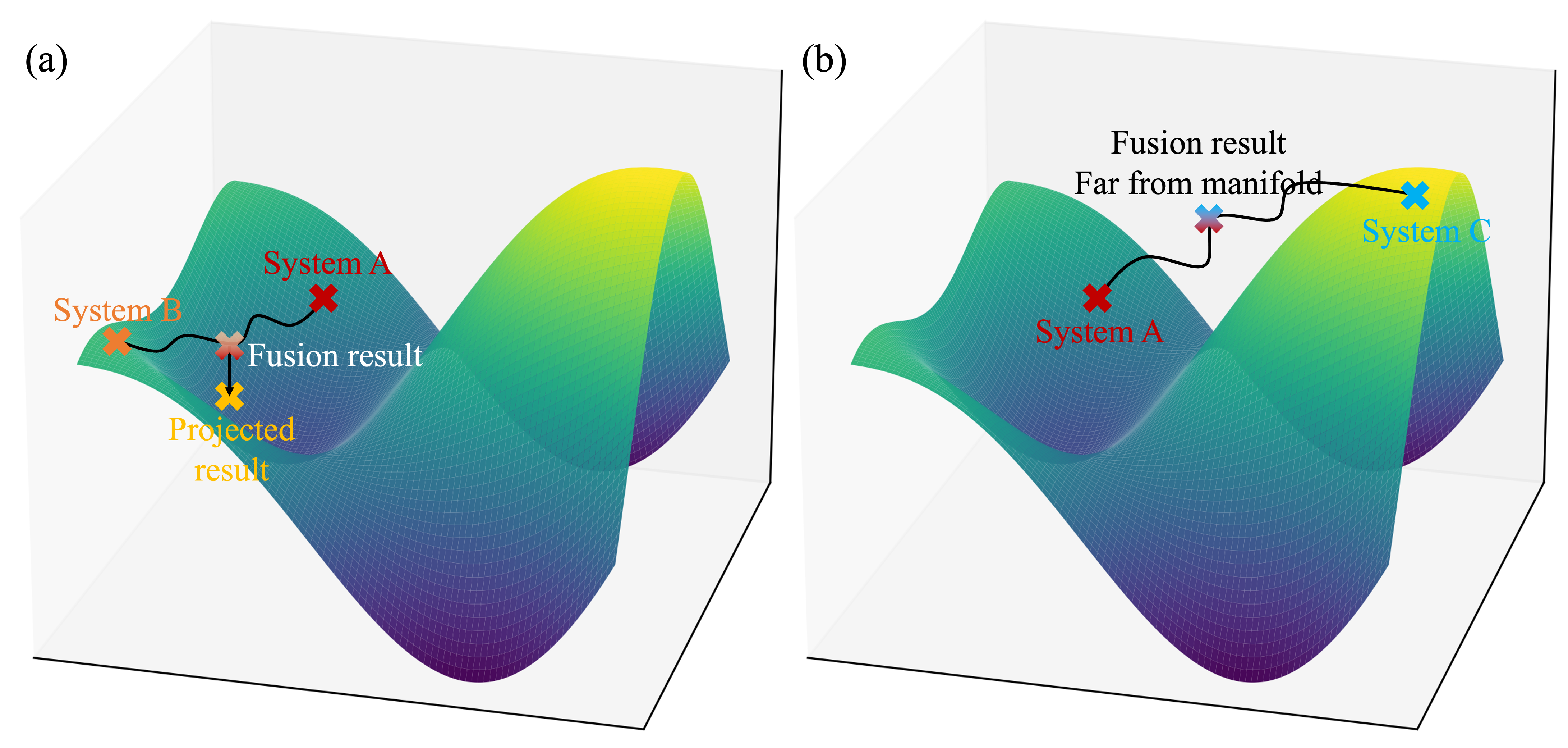}
\caption{Schematic of feature-space fusion from a manifold perspective. (a) Fusion between relatively similar parent systems. (b) Fusion between dissimilar parent systems. The illustration is conceptual and does not represent an explicitly identified latent manifold.}
\label{manifold}
\end{figure}

\codexnew{Within this picture, fusion between relatively similar parents may remain closer to the region represented by the training data, whereas fusion between dissimilar parents may leave that region and yield trivial dynamics or finite-time blow-up. The two possibilities are illustrated schematically in Fig.~\ref{manifold}. They offer a qualitative way to describe outcomes observed in the fusion experiments, rather than a formal geometric criterion for system similarity.}
 \FloatBarrier
\section{\codexnew{Dissimilarity Matrices and Spectral Embeddings}}

\begin{figure}[!h]
\centering
\includegraphics[width=\columnwidth]{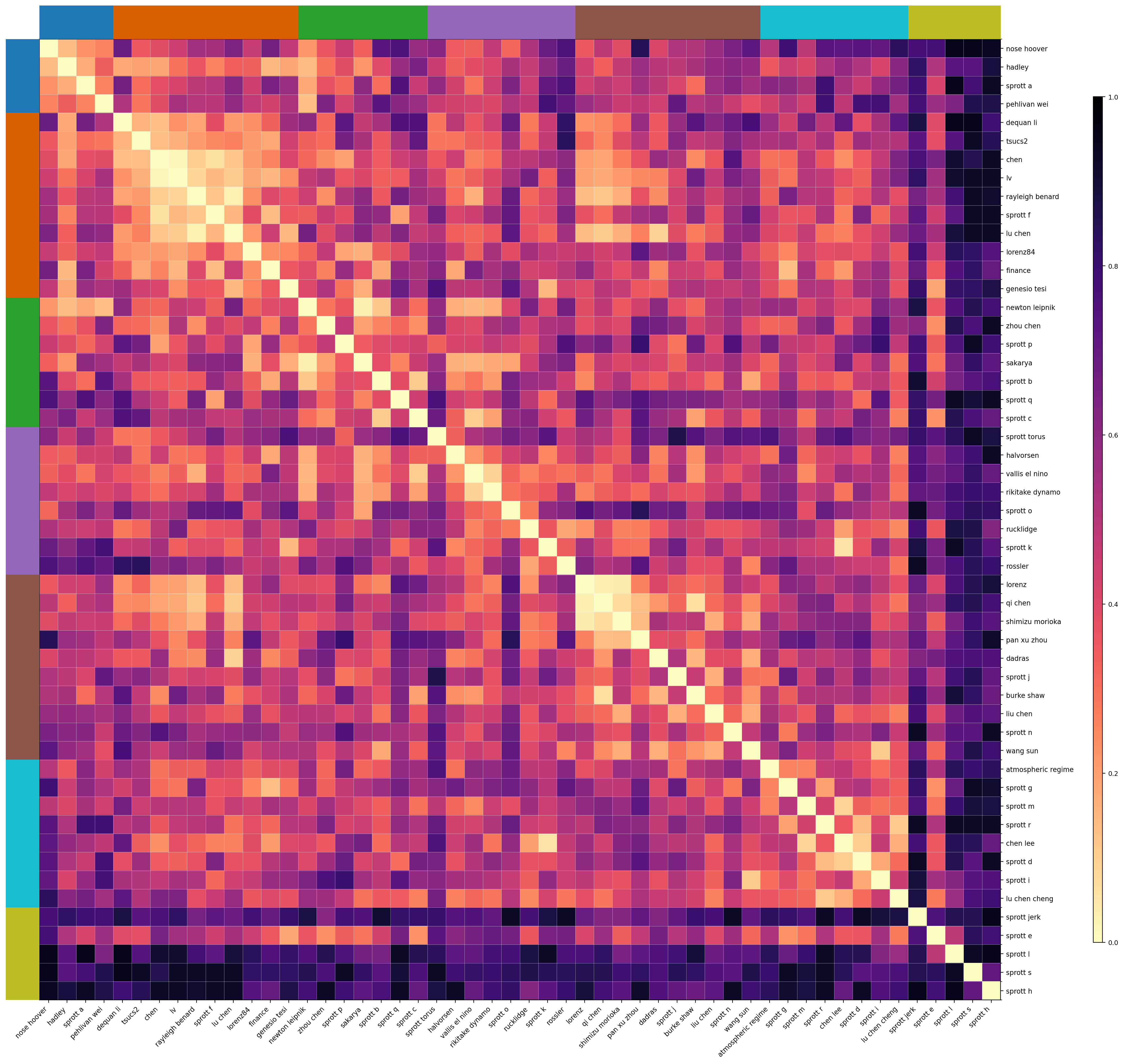}
\caption{\codexnew{Reordered NDTL-induced dissimilarity matrix for the 52 retained systems. The ordering and color strip indicate the spectral clustering labels obtained after applying a radial basis function kernel to the pairwise dissimilarities.}}
\label{fig:classification52_matrix}
\end{figure}

\begin{figure}[!h]
\centering
\includegraphics[width=0.48\textwidth]{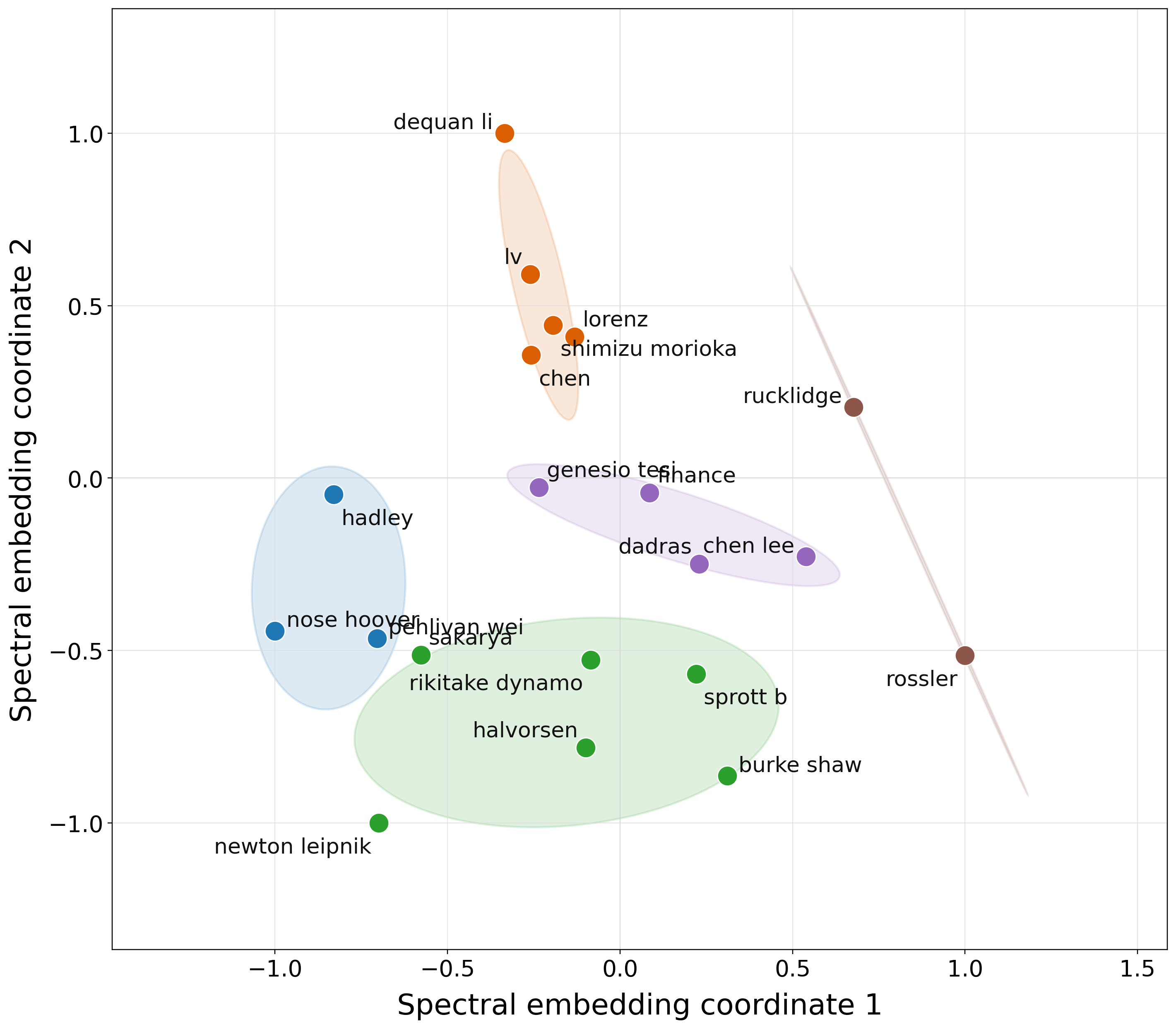}
\caption{Spectral clustering of a 20-system subset using the NDTL-induced distance. This visualization shows that the distance obtained from content feature fusion organizes canonical chaotic systems into coherent groups.}
\label{fig:classification20_spectral}
\end{figure}

\begin{figure}[!h]
\centering
\includegraphics[width=\columnwidth]{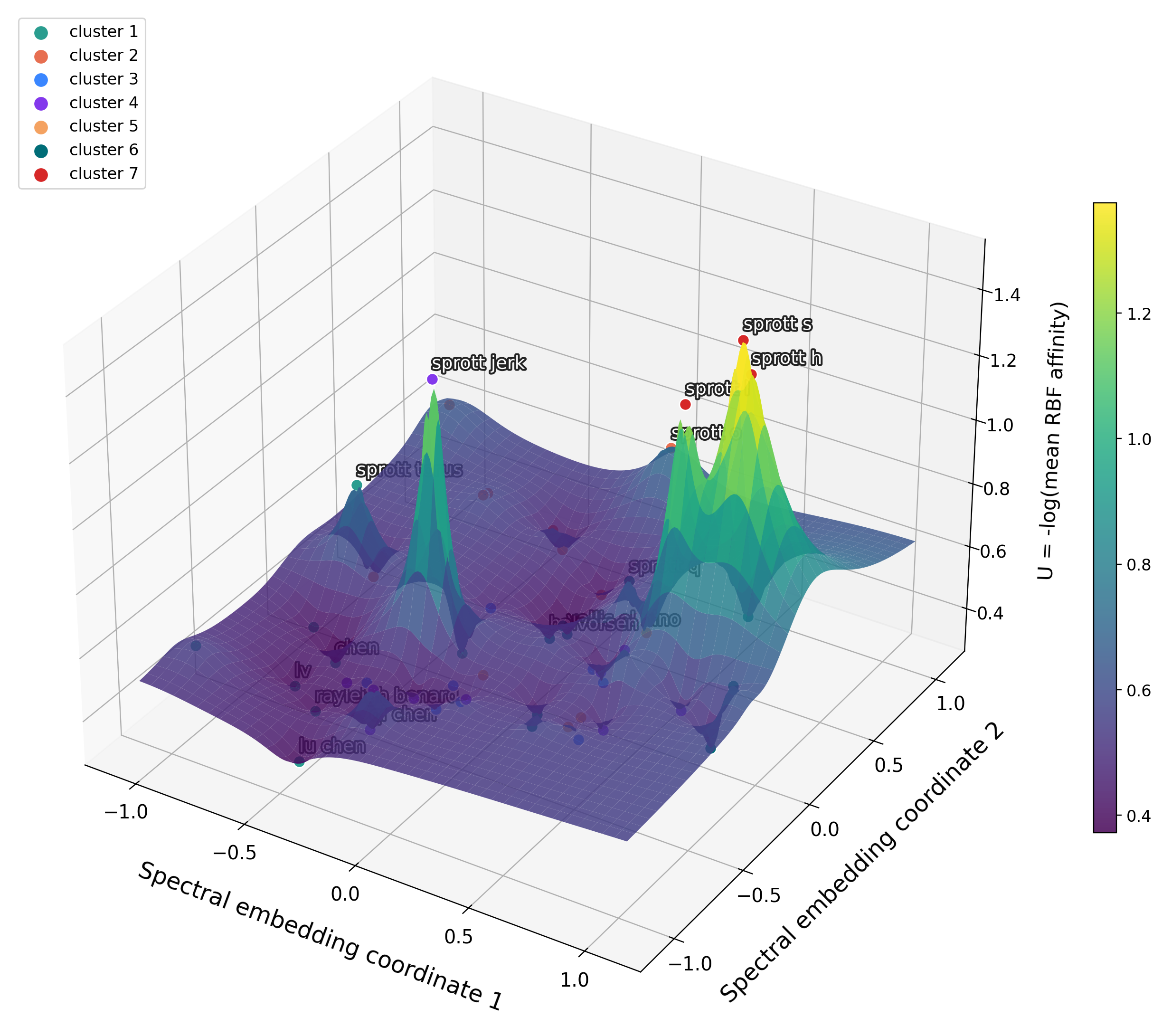}
\caption{Three-dimensional visualization of the spectral embedding shown in the main text. The surface height is \(U=-\log\) of the mean radial basis function affinity, and colors denote the seven spectral clusters.}
\label{fig:classification52_surface}
\end{figure}

\codexnew{The reordered distance matrix in Fig.~\ref{fig:classification52_matrix} displays the pairwise structure used by the clustering procedure. Blocks of lower dissimilarity correspond to groups whose members admit comparatively continuous fusion paths, whereas larger intergroup values indicate stronger disruption of the generated dynamics. Figure~\ref{fig:classification20_spectral} gives a direct visualization on a 20-system subset, while Fig.~\ref{fig:classification52_surface} provides a complementary three-dimensional view of the full spectral embedding reported in the main text.}

\section{\codexnew{Ecological Creation in the Hastings-Powell Food Chain Model}}
\label{app:ecology}
\label{sec:hp_holling_creation}

This section provides the technical details for the Hastings Powell (HP) food chain experiment reported in the main text. The goal is to test whether NDTL can create a new persistent three species ecological module from two parent food chain systems with distinct long time dynamics.

\subsubsection{HP model family}
We consider a three species food chain consisting of a plant or resource species \(P\), an herbivore \(H\), and a carnivore \(C\). In ecological coordinates, the vector field is written as
\begin{equation}
\label{eq:hp_holling_sm}
\begin{aligned}
\dot P &= P\left(1-\frac{P}{K}\right)-u_h(P)H,\\
\dot H &= H\left(u_h(P)-d_h\right)-u_c(H)C,\\
\dot C &= C\left(u_c(H)-d_c\right),
\end{aligned}
\end{equation}
where the two saturating uptake functions are
\begin{equation}
\label{eq:hp_uptake_sm}
u_h(P)=\frac{\alpha_h\beta_h P}{\gamma_h+P},
\qquad
u_c(H)=\frac{\alpha_c\beta_c H}{\gamma_c+H}.
\end{equation}
In the parameterization used here, \(\beta_h=\beta_c=1\). The original HP notation writes the herbivore uptake as \(a_1P/(1+b_1P)\), which corresponds to \(\gamma_h=1/b_1\) and \(\alpha_h=a_1/b_1\) in Eq.~\eqref{eq:hp_uptake_sm}. With \(a_1=5\), the parent systems used in the experiments are listed in Table~\ref{tab:hp_parents_sm}.

\begin{table}[h!]
\centering
\caption{HP parent systems used in the ecological system creation experiment.}
\label{tab:hp_parents_sm}
\begin{ruledtabular}
\begin{tabular}{lccccccc}
Parent & \(b_1\) & \(\alpha_h\) & \(\alpha_c\) & \(\gamma_h\) & \(\gamma_c\) & \(d_h\) & \(d_c\) \\
\hline
Stable   & 2.0 & 2.500 & 0.050 & 0.500 & 0.500 & 0.400 & 0.010 \\
Periodic & 2.3 & 2.174 & 0.050 & 0.435 & 0.500 & 0.400 & 0.010 \\
Chaotic  & 3.0 & 1.667 & 0.050 & 0.333 & 0.500 & 0.400 & 0.010 \\
\end{tabular}
\end{ruledtabular}
\end{table}

\subsubsection{Training and fusion protocol}
The main HP experiments use the fixed ecological domain
\[
    [0,2]\times [0,2]\times [0,15].
\]
Each parent vector field is sampled on a \(64^3\) grid. For each parent pair, we train a separate CAE rather than reusing a checkpoint trained on Lorenz like systems. The training set is augmented by parent vector field samples, \(10\%\) multiplicative parameter perturbations, and Gaussian tensor noise with standard deviation \(0.002\). Content feature fusion is performed at
\[
    \alpha_b\in\{0.00,0.25,0.50,0.75,1.00\},
\]
where \(\alpha_b=0\) and \(\alpha_b=1\) correspond to the two parent endpoints. The fused tensor field is evaluated by trilinear interpolation in ecological coordinates.

To obtain a mechanistically interpretable ecological module, we project the fused vector field back onto the HP model family in Eqs.~\eqref{eq:hp_holling_sm} and \eqref{eq:hp_uptake_sm}. This projection is obtained by fitting the saturation profiles: the outer search is performed over \((\gamma_h,\gamma_c)\), and the remaining positive coefficients are estimated by constrained least squares. We report both the raw fused NDTL tensor field and the projected system in the HP model family, but the main text ecological conclusion is based on the projected system.

\subsubsection{Stable to chaotic fusion}
The stable to chaotic experiment fuses the stable HP parent with the chaotic HP parent. Table~\ref{tab:hp_stable_chaotic_sm} summarizes the single initial condition evaluation. The midpoint \(\alpha_b=0.50\) is the main ecological creation result: it remains persistent, exhibits a large number of carnivore peaks in the single initial condition test, and can be projected back to the HP model family with relative root mean square error (relative RMSE) \(7.434\times 10^{-2}\). In the table, Pers. denotes persistence of all three species, \(\lambda_{\max}\) is the largest Lyapunov exponent, Peaks is the number of detected carnivore peaks, F. loss is the final fusion loss, HP RMSE is the relative RMSE after projection to the HP model family, and basis RMSE is the relative RMSE obtained with the full basis used in the projection fit.

\begin{table}[h!]
\centering
\caption{Stable to chaotic HP fusion results on the fixed ecological domain.}
\label{tab:hp_stable_chaotic_sm}
\resizebox{\columnwidth}{!}{
\begin{tabular}{ccccccccc}
\hline
\(\alpha_b\) &
Pers. &
\(\lambda_{\max}\) &
Peaks &
F. loss &
HP RMSE &
Basis RMSE &
\(\hat{\alpha}_h\) &
\(\hat{\gamma}_h\) \\
\hline
0.00 & \(\checkmark\) & \(-4.907{\times}10^{-2}\) & 0   & 0                         & \(3.245{\times}10^{-8}\) & \(3.002{\times}10^{-8}\) & 2.500 & 0.500 \\
0.25 & \(\checkmark\) & \(-3.865{\times}10^{-1}\) & 0   & \(2.778{\times}10^{-1}\) & \(5.280{\times}10^{-2}\) & \(5.275{\times}10^{-2}\) & 2.288 & 0.464 \\
0.50 & \(\checkmark\) & \(-1.376{\times}10^{-2}\) & 467 & \(3.705{\times}10^{-1}\) & \(7.434{\times}10^{-2}\) & \(7.429{\times}10^{-2}\) & 2.077 & 0.423 \\
0.75 & \(\checkmark\) & \(-4.518{\times}10^{-1}\) & 0   & \(2.784{\times}10^{-1}\) & \(1.107{\times}10^{-1}\) & \(1.105{\times}10^{-1}\) & 1.864 & 0.377 \\
1.00 & \(\checkmark\) & \( 1.090{\times}10^{-2}\) & 7   & 0                         & \(3.732{\times}10^{-7}\) & \(1.850{\times}10^{-7}\) & 1.667 & 0.333 \\
\hline
\end{tabular}
}
\end{table}

For the \(\alpha_b=0.50\) projected HP system, the fitted parameters are
\begin{equation*}
\begin{aligned}
\hat K &= 0.9992, &
\hat{\alpha}_h &= 2.0766,\\
\hat{\alpha}_c &= 0.0510, &
\hat{\gamma}_h &= 0.4234,\\
\hat{\gamma}_c &= 0.5355, &
\hat d_h &= 0.3991,\\
\hat d_c &= 0.00968.
\end{aligned}
\end{equation*}

Thus the effective HP saturation parameter is
\[
    b_1^{\mathrm{eff}}
    =
    \frac{1}{\hat{\gamma}_h}
    \approx
    2.36,
\]
which lies between the stable parent \(b_1=2.0\) and the chaotic parent \(b_1=3.0\).

\subsubsection{Multiple initial condition validation}
We validate the \(\alpha_b=0.50\) stable to chaotic case using 27 initial conditions,
\[
\begin{gathered}
P_0\in\{0.2,0.5,0.9\},\\
H_0\in\{0.05,0.18,0.36\},\\
C_0\in\{7.5,9.5,13.0\}.
\end{gathered}
\]

Each trajectory is integrated with \(dt=0.05\), \(90{,}000\) total steps, and \(20{,}000\) kept steps after transient removal. The results are summarized in Table~\ref{tab:hp_multistart_sm}.
In this table, Pers. and Osc. give the numbers of persistent and oscillatory trajectories among the 27 initial conditions. Med. denotes the median over initial conditions, std. is the standard deviation of the retained trajectory norm, and CV denotes the coefficient of variation of carnivore peak intervals. Across CV is the coefficient of variation of the mean peak interval across initial conditions.

\begin{table}[h!]
\centering
\caption{Multiple initial condition validation for the stable to chaotic \(\alpha_b=0.50\) case.}
\label{tab:hp_multistart_sm}
\resizebox{\columnwidth}{!}{
\begin{tabular}{lcccccc}
\hline
System &
Pers. &
Osc. &
Med. peaks &
Med. std. &
Med. CV &
Across CV \\
\hline
Stable exact &
27/27 & 0/27  & 0  & \(3.413{\times}10^{-12}\) & n/a & n/a \\
Chaotic exact &
27/27 & 27/27 & 22 & \(9.044{\times}10^{-1}\) & \(3.910{\times}10^{-1}\) & \(2.873{\times}10^{-2}\) \\
Raw NDTL &
23/27 & 7/27  & 0  & n/a                        & \(5.003{\times}10^{-1}\) & \(2.153{\times}10^{-1}\) \\
Projected HP &
27/27 & 27/27 & 15 & \(2.706{\times}10^{-1}\) & \(3.839{\times}10^{-4}\) & \(2.543{\times}10^{-5}\) \\
\hline
\end{tabular}
}
\end{table}

The raw fused NDTL tensor field is persistent for most, but not all, tested initial conditions. In contrast, the projected HP system is persistent and oscillatory for all 27 initial conditions, with a very small across initial condition peak interval coefficient of variation. This is why the main text emphasizes the projected system in the HP model family as the interpretable ecological creation result.

\subsubsection{Floquet diagnostic}
For the \(\alpha_b=0.50\) projected HP system, we further perform a numerical Floquet check. The estimated period is
\[
    T = 65.1349,
\]
with closure norm
\[
    3.438\times 10^{-4}.
\]
The absolute Floquet multipliers are
\[
    \left[
    0.9444,\;
    1.425\times 10^{-13},\;
    0.9984
    \right].
\]
The multiplier close to one corresponds to the phase direction of the periodic orbit, and the remaining transverse multipliers are less than one in magnitude. This supports the interpretation of the projected \(\alpha_b=0.50\) HP system as a transversely stable ecological limit cycle.
 \FloatBarrier
\section{\codexnew{Chaotic Image Encryption Benchmark}}
\label{app:encryption}
\label{sec:crypto_benchmark}

To validate the usability of the generated dynamics in downstream applications, we embed the fused system generated by combining the content of the Lorenz system with the style of Chen's system into a standard permutation diffusion image encryption pipeline. In this pipeline, permutation reorders pixel positions, and diffusion changes pixel intensities using a mask generated from a chaotic trajectory. For comparison, the parent systems Lorenz and Chen are also evaluated under the same protocol. The encryption pipeline performs image permutation followed by two pass diffusion using an 8 bit mask generated from the chaotic trajectory. All systems are tested on three standard images: \textit{camera}, \textit{moon}, and \textit{astronaut} with size \(512 \times 512\). The initial condition for integration is set to \([0.3,0.3,0.5]\), with \(dt = 0.01\) and burn in 1000 steps. Key perturbation is \(1 \times 10^{-14}\) for sensitivity testing.

\subsection{Benchmark metrics}
Each system is evaluated on the following metrics:

\begin{itemize}
    \item \textbf{Cipher entropy} measures the information content of the ciphertext.
    \item \textbf{Adjacent pixel correlation} measures residual correlation among neighboring pixels.
    \item \textbf{NPCR and UACI} denote the number of pixels change rate and the unified average changing intensity, respectively, and quantify sensitivity to plaintext changes.
    \item \textbf{Key NPCR and Key UACI} apply the same two metrics after perturbing the initial condition.
    \item \textbf{Aggregate score} is computed as the mean of normalized metrics.
\end{itemize}

\subsection{Results}
Table~\ref{tab:crypto_full} summarizes the benchmark results for the selected fused system and its parent systems. The fused system achieves the highest aggregate score among the three, while Lorenz exhibits lowest adjacent pixel correlation and Chen performs well on selected images. This table provides a complete comparison of system performance under the full encryption protocol.

\begin{table}[h]
\centering
\caption{Chaotic image encryption benchmark for Lorenz, Chen, and the selected NDTL generated fused system.}
\label{tab:crypto_full}
\resizebox{\columnwidth}{!}{%
\begin{tabular}{lccccccc}
\hline
System &
Entropy &
Corr. &
NPCR &
UACI &
Key NPCR &
Key UACI &
Score \\
\hline
Lorenz & 7.999235 & 0.001060 & 99.607 & 33.473 & 99.605 & 33.401 & 0.999407 \\
Chen & 7.999248 & 0.002044 & 99.595 & 33.485 & 99.600 & 33.498 & 0.999223 \\
Fused & 7.999325 & 0.001306 & 99.613 & 33.448 & 99.614 & 33.475 & 0.999484 \\
\hline
\end{tabular}
}
\end{table}

All systems produce visually noise like ciphertext and allow exact reconstruction with the correct key. The selected fused system therefore provides a usable chaotic source in the same encryption pipeline as the parent systems.
 
\bibliography{prr}
\end{document}